\documentclass[usedcolumn,usenatbib]{mn2e}
\usepackage{graphicx}
\usepackage{times}
\usepackage{dcolumn}
\newlength{\colwidth}
\newlength{\plotwidth}
\newlength{\twothirdwidth}
\newlength{\reallyfullheight}
\newcommand{\ha}{H$\alpha$ }
\newcommand{\hb}{H$\beta$ }
\newcommand{\hg}{H$\gamma$ }

\newcommand{\oiii}{[O\,III] }
\newcommand{\oii}{[O\,II] }

\newcommand{\nii}{[N\,II] }

\title[Spatial decomposition of on-nucleus spectra of quasar host
 galaxies]{Spatial decomposition of on-nucleus spectra of quasar host
 galaxies\thanks{Based on observations made with telescopes at the European
   Southern Observatories La Silla and Paranal Observatories in Chile
   under programme IDs 60.B-0320 and 65.P-0361.}
}

\author[K.~Jahnke et al.]{K.~Jahnke$^1$, L.~Wisotzki$^2$, F.~Courbin$^3$,
	G.~Letawe$^4$\\ 
	$^1$Max-Planck-Institut f\"ur Astronomie, K\"onigstuhl 17, 69117
 Heidelberg, Germany\\
	$^2$Astrophysikalisches Institut Potsdam, An der Sternwarte
16, 14482 Potsdam, Germany\\
        $^3$Laboratoire d'Astrophysique, Ecole Polytechnique F\'ed\'eral de
	Lausanne (EPFL), Observatoire, CH-1290 Sauverny, Switzerland\\
        $^4$Institut d'Astrophysique et G\'eophysique, Universit\'e de
	L\`iege, All\'ee du 6 Aout, 17, Sart Tilman (Bat. B5C), B-4000
	L\`iege, Belgium
}
\pagerange{\pageref{firstpage}--\pageref{lastpage}}
\pubyear{2007}

\begin{document}

\maketitle

\label{firstpage}

\begin{abstract}
In order to study the host galaxies of type~1 (broad-line) quasars, we present
a semi-analytic modelling method to decompose the on-nucleus spectra of
quasars into nuclear and host galaxy channels. The method uses the spatial
information contained in long-slit or slitlet spectra. A routine determines
the best fitting combination of the spatial distribution of the point like
nucleus and extended host galaxy. This is fully complementary to a numerical
spatial deconvolution technique that we applied to the same data in a previous
analysis, which allows a cross-calibration of the two methods. Inputs are a
simultaneously observed PSF, and external constraints on galaxy morphology
from imaging. We demonstrate the capabilities of the method to two samples of
a total of 18 quasars observed with EFOSC at the ESO 3.6m telescope and FORS1
at the ESO VLT.
$\sim$50\% of the host galaxies with sucessful decomposition show distortions
in their rotation curves or peculiar gas velocities above normal maximum
velocities for disks. This is consistent with the fraction from optical
imaging. All host galaxies have quite young stellar populations, typically
1--2~Gyr. For the disk dominated hosts these are consistent with their
inactive counterparts, the luminosity weighted stellar ages are much younger
for the bulge dominated hosts, compared to inactive early type galaxies. While
this presents further evidence for a connection of galaxy interaction and AGN
activity for half of the sample, this is not clear for the other half: These
are often undistorted disk dominated host galaxies, and interaction on a
smaller level might be detected in deeper high-resolution images or deeper
spectroscopic data. The velocity information does not show obvious signs for
large scale outflows triggered by AGN feedback -- the data is consistent with
velocity fields created by galaxy interaction.
\end{abstract}


\section{Introduction}\label{sec:spec}
After having been a fascinating but bizarre species of astrophysical objects
for three decades, quasars have shifted into the mainstream focus of galaxy
evolution. When the tight relation between black hole masses and masses of the
galaxies hosting them became evident \citep[e.g.,][]{ferr00} it was clear that
there is a mechanism leading to an effective co-evolution of galaxies and
their central black holes. Quasars, accreting black holes, might be the key to
understanding the extent and physics of this co-evolution, or at least of the
phase in which the black holes grow strongest.

In a second role, the energy feedback of quasars via galactic winds currently
provides an alternative possibility beside supernova feedback to expell gas
from galaxies and quench star formation. Quasar activity might be responsible
for the switching off of star formation and the onset of evolution of massive
galaxies towards the red sequence of ellipticals.

For both roles vital information is still missing before the question about
cause and effect in co-evolution can be settled, the same is true for the
frequency of outflows in quasars, their properties, and the effect on the
interstellar medium (ISM) in the host galaxies. In the end both roles might be
the two sides of the same complex mechanism.

Most studies of quasar host galaxies in the past were restricted to
imaging. For investigating e.g.\ morphologies, companion statistics,
interaction states, luminosity distributions, imaging is well suited. High
resolution imaging with ground based telescopes or the HST
\citep[e.g.,][]{bahc97,sanc04a,jahn04b,floy04} provided most of todays
knowledge on quasar host galaxies.

However, many techniques applied to the investigation of normal, inactive
galaxies require SED information in the shape of colours or even
spectra. Determination of e.g.\ stellar populations, velocity measurements,
ionisation states of the ISM can, by their nature, not be performed on imaging
data. So spectral information is not only appreciated, but in the face of
fundamental questions as `Is nuclear activity part of any galaxie's life
cycle?', `What triggers nuclear activity?'  or `How far does co-evolution go?'
it is urgently needed.
\smallskip

In a recent article \citep{jahn04a} we described a method to extract coarse
but robust spectral information on QSO host galaxies. Coarse, because by using
broad band imaging we sample the SED of a QSO in our case in seven wavelength
intervals between 4000~\AA\ and 23000~\AA. The broad band filters integrate
over several hundred or thousand \AA, in order to collect significant amounts
of light.

The method showed to be robust, because by now the uncertainties and caveats
of semi-analytic two-dimensional image modelling are quite well
understood. Software has been developed and tested, by us \citep{kuhl04} and
other groups \citep[e.g.,][]{peng02,floy04}, and the problems of PSF and host
modelling have been investigated in depth
\citep[e.g.,][]{kuhl04,jahn04b,sanc04a,jahn06b}. Thus the implemented
modelling scheme is powerful and has assessable uncertainties, as was
demonstrated in several articles.

In spectroscopy the situation is different. Spectroscopic studies of the host
galaxies of broad-line, type~1 QSOs are very difficult: The AGN in the centre
overpowers the host galaxy's emission much more strongly than in broadband
imaging, because the narrow slit is integrating much of a compact nucleus and
little of an extended galaxy. Until recently it was not possible to access
host spectra directly around the nucleus at all, where the connection between
host and nucleus is strongest.

\subsection{Previous studies and fundamental problems}
Very few spectroscopic studies were attempted, knowing too well of the
problems of the task. In the 30 years of host galaxy research before 2006 only
about 25 articles were published on quasar host galaxy spectroscopy --
compared to more than 100 on imaging --, and more than half of those appeared
before 1985.

Between the mid 70's and mid 80's pioneering spectroscopic work was done,
confirming the cosmological nature of the QSO redshift
\citep{wamp75,stoc76,rich77,wyck80a}, and establishing the presence of gas
emission lines and later stellar continuum and absorption lines underneath the
nuclear glare \citep{mort78,gree78,wyck80b,boro82a,mack84}. 

The first larger sample was investigated in a series of papers by
\citet*{boro82}, \citet*{boro84} and \citet*{boro85}, which also remained the
largest study for almost 20 years. Two characteristic groups of hosts were
classified there, with either blue continua and strong gas lines, or continua
with weak or absent gas emission. The authors found positive correlations of
this classification with the shape of the nuclear Balmer lines, the strength
of the nuclear \oiii emission, the strength of Fe features, absorption line
strength, radio spectrum shape and radio morphology. Later studies found
evidence for star formation possibly induced by tidal events
\citep{mack84,hutc88}, coordinated rotation of gas in the host
\citep{hick87,hutc90}, and an extention of the gas emission beyond the
continuum \citep{hutc88}.

The next level of detail took ten more years when \citet{cana00a} performed a
deep, spatially resolved study of the host galaxy of 3C~48. The analysis
included modelling the stellar content of 32 regions of the host. They found
strong evidence for recent starburst of 5 to 100~Myrs and old evolved stars in
other parts, fitting well into the picture of a merger event going on in
3C~48. This result is supported by -- or rather vice versa the study was
initiated because of -- the tidally disturbed structure of the object visible
in imaging.

At the same time a study aimed to compare the stellar populations of
radio-quiet QSOs, radio-loud QSOs and radio galaxies \citep{hugh00,nola01} --
the objects investigated were part of an earlier well studied imaging sample
\citep{dunl93,tayl96,dunl03}. Model fitting showed evolved stellar
populations, with only small contributions to the light from recent
star-formation, corresponding to young stars of less then about one percent of
the visible mass. The samples however, were manually selected to match in
different properties, as luminosity and redshift range, and contained
extremely massive galaxies, thus it is difficult to assess in which properties
these are representative for the general host galaxy population.

In the recent past, a number of publications used the wealth of the Sloan
Digital Sky Survey (SDSS) as an AGN database for host galaxy studies. While
most of them focussed on type~2 quasars or rather Seyferts, particularly to
mention the large study of Sy~2 by \citet{kauf03} and of type~2 QSOs
\citep{zaka03,zaka05}, only recently studies \citep{hao05,dong05,vand06}
tackled SDSS broad-line AGN, using an spectral eigenvector decomposition
technique. Type~2 AGN have, from their lack of a directly visible central
source, no contamination problem in the first place. However, they only tell
half of the story: The central source can not be studied directly, black hole
masses can only be estimated indirectly from the velocity dispersion or
luminosity of their host galaxy -- without the possibility of an independent
calibration. In addition, the obscuring material might be different or
differently distributed at different luminosities. Particularly, few proper
type~2 quasars are known. Until the unified model of type~1 and 2 is
confirmed, and also to confirm it, studies of the more difficult to analyze
type~1 QSOs are required.
\medskip

Aside from the mentioned studies from the SDSS, in all other work on type~1
AGN the host spectra were either taken with the slit directly ``on-nucleus'',
but were strongly contaminated by nuclear light, or taken ``off-nucleus'', at
distances of several arcseconds from the centre, to sample regions of the host
as far away as possible from the nuclear light.

Two problems arise from this: Firstly, a distance of 5\arcsec as e.g.\ used by
\citet{hugh00} corresponds to a linear distance of $\sim$20~kpc at $z=0.2$ --
as a consequence, the slit does not sample the central regions of the host
galaxy. Secondly, despite the 2--5\arcsec distance of the slits from the
nucleus there remains a nuclear contribution, scattered into the slit by the
seeing. These nuclear contributions were in the abovementioned studies either
not corrected or it was attempted to remove them by subtracting a scaled
on-nucleus spectrum that was assumed to contain only nuclear light. As in the
case of one-dimensional spatial host galaxy image separation, the amount of
nuclear light to be subtracted is not known a priory. The only knowledge lies
in that the stellar continuum does not contain broad emission lines. So
increasing amounts of on-nucleus spectrum were subtracted, modified with an
ad-hoc function of wavelength, until the broad lines (typically \ha and/or
\hb) were removed. This method is often imperfect and can not remove the
contribution completely and leaves unwanted residual flux
\citep[e.g.][]{hutc90}.

In summary, with off-nucleus spectra neither can the astrophysically most
interesting central parts of host galaxies be sampled, nor does this approach
guarantee the absence of artefacts like nuclear residuals in the resulting
spectra.

\subsection{On-nucleus data: the test cases}
To avoid these drawbacks, we acquired \textit{on-nucleus} slit spectroscopy
data for two samples of broad-line quasars at two different telescopes and
instruments, EFOSC at the ESO 3.6m telescope and FORS1 at the ESO VLT (sample
summary in Tab.~\ref{tab:sampleprops}). Such on-nucleus data require, before
analysis, that the contributions of nucleus and host galaxy have to be
separated in the data processing stage to access the host galaxy
information. In order to do this, we developed two complementary
techniques. In this article we present for one of the techniques both the
working principles, as well as summarize the most important science results
from these two datasets. The other, complementary deconvolution technique was
applied to data from the same VLT-campaign (extended by a second semester,
doubling the number the ten objects analysed here to 20), and we published the
results in more detail in a series of other papers \citep[see
below,]{cour02b,leta04,maga05,leta06}.

The extraction of the host galaxy spectrum is made possible by these two
independent and new techniques that allow to separate the host from nuclear
light, making use of the so far unused spatial dimension of long-slit
spectra. These methods allow to estimate the host SED directly from on-nuclear
long slit spectra, without prior assumptions about the resulting host
\textit{spectrum}, and allow a spatially resolved investigation directly down
to the centre.

The first method is the spatial numerical deconvolution\footnote{Throughout
the paper we use the term ``deconvolution'' in the mathematical sense,
contrary to some authors, who use ``deconvolution'' to denote any method
applied to remove the nuclear contribution. For this we use the more
appropriate term ``decomposition''.}  of the two-dimensional (2d) composite
spectrum developed by \citet{cour00}, already mentioned above. The other is a
simultaneous semi-analytic spatial modelling of nucleus and host spectrum,
which we describe in the following. Both methods are based on the same
principles -- using the spatial information contained in 2d spectra -- yet the
algorithmic approach is different and complementary. \citeauthor{cour00} are
deconvolving the spatial component of the spectrum into a point source and an
extended source, using a modified version of their algorithms for imaging
deconvolution \citep{maga98}. Using the two techniques on the the same VLT
dataset, allows a comparison of the techniques for identical input data.
\medskip

Our previous work focused on spatial deconvolution of the data. We are now
investigating a fundamentally different approach, with semi-analytic
two-component spatial PSF fitting of the 2d-spectra. In the following we will
describe the general principles of the method and detailed implementation. In
section~\ref{sec:applic} we present the application of this method to one
example object for each of the two different qualities of spectroscopic data
taken with ESO 3.6m EFOSC and VLT FORS. In section~\ref{sec:results} we
summarize the scientific results we extracted from the two datasets.

A discussion of the capabilities and limits of the modelling technique
is given in section~\ref{sec:discussion}.

\section{2d decomposition of spectra}\label{sec:spec2d}
It is possible to numerically disentangle host galaxies from the active
nucleus. For spectroscopy this is more difficult than for imaging, because of
the additional parameter wavelength $\lambda$, and less S/N at each $\lambda$
point than in broadband imaging. Our approach is to incorporate as much
external information into the separation process as possible. ``Information''
in this case means knowledge about the QSO, like morphological host galaxy
parameters, the shape of the PSF or other information from imaging. We combine
this with using all {\it spatial} information available in 2d longslit
spectra. Up to now this information was never used for the extraction of host
galaxy spectra.

\subsection{Different instruments and data types}\label{sec:datatypes}
Optical and NIR spectral information can be obtained with a multitude of
instruments and techniques. Beside classical long slit spectroscopy, currently
several types of multi-object spectrographs are available. Units like the
multi object spectroscopy unit (MOS) of FORS on the ESO VLT use movable
slitlets of fixed length, or use exchangeable masks that are individually
manufactured for each observation (e.g.\ at the ESO VLT for the FORS2 and
VIMOS instruments). Slitlets and masks in general have the same capabilities
as long slits, with an additional degree of freedom when targetting more than
one object in the field of view.

All of these spectroscopy techniques have in common that spectral and spatial
information are obtained simultaneously. For host galaxy applications the
spatial information can be used to disentangle the spectral components of an
active nucleus and host galaxy. The approach described below can be used with
long slit, slitlet, or mask spectra \citep[we have recently also modified this
principle to integral field spectroscopy, see][]{sanc06}.

\subsection{General approach}\label{sec:spec2dintro}

\begin{figure*}
\begin{center}
\begin{minipage}[b][8.5cm][t]{7cm}
\includegraphics[bb = 10 15 310 394,clip,angle=0,width=\colwidth]{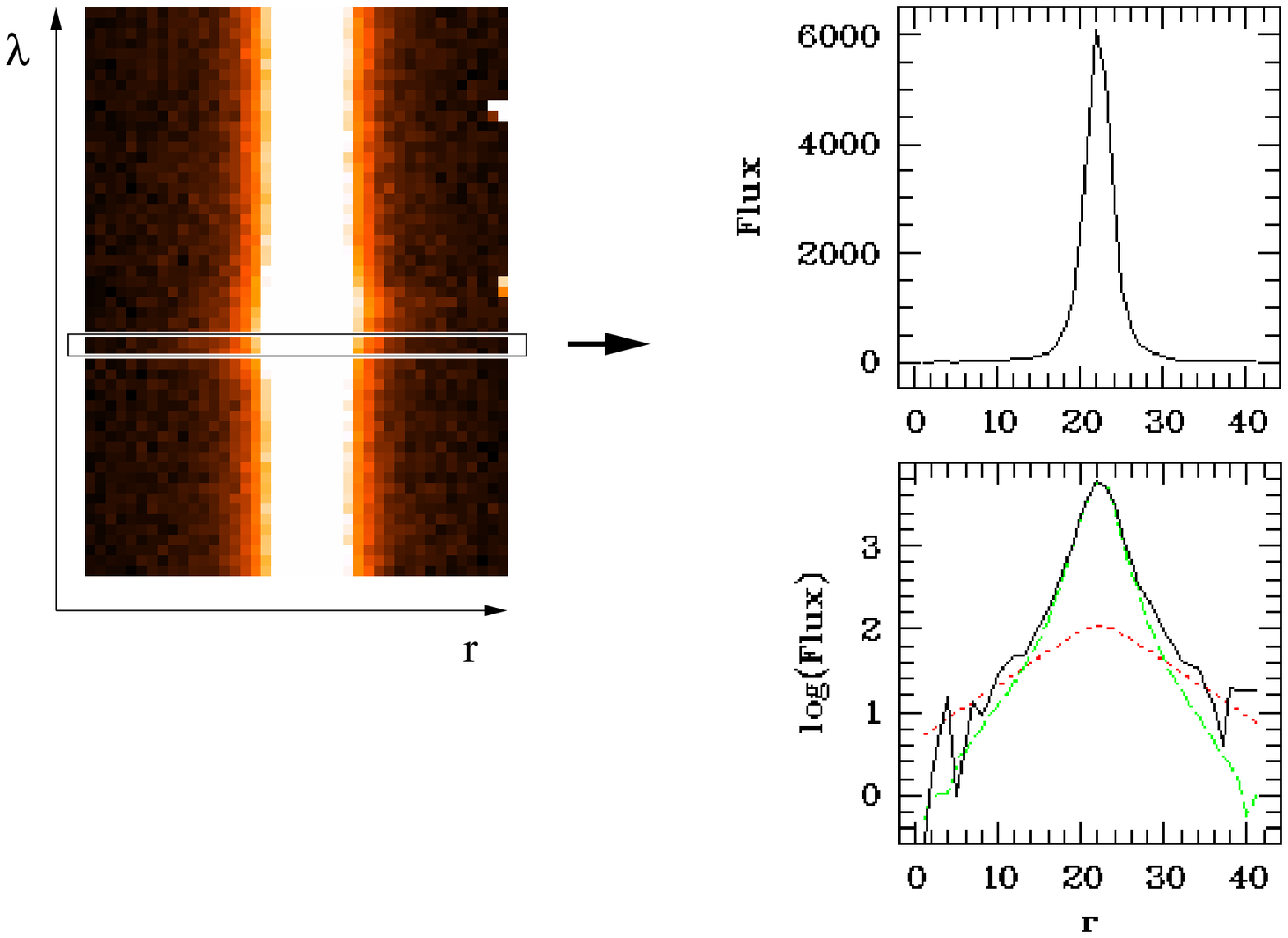}
\end{minipage}
\begin{minipage}[b][8.5cm][t]{6cm}
\includegraphics[bb = 55 598 252 772,clip,angle=0,width=5cm]{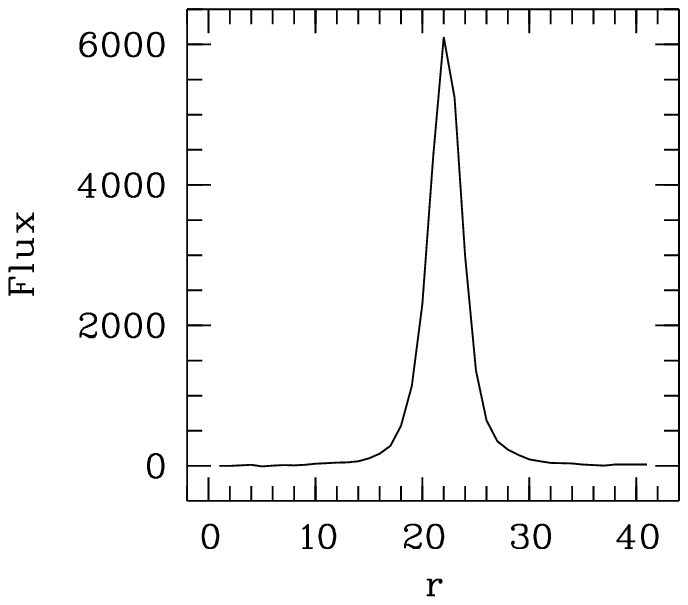}\\
\includegraphics[bb = 55 598 252 772,clip,angle=0,width=5cm]{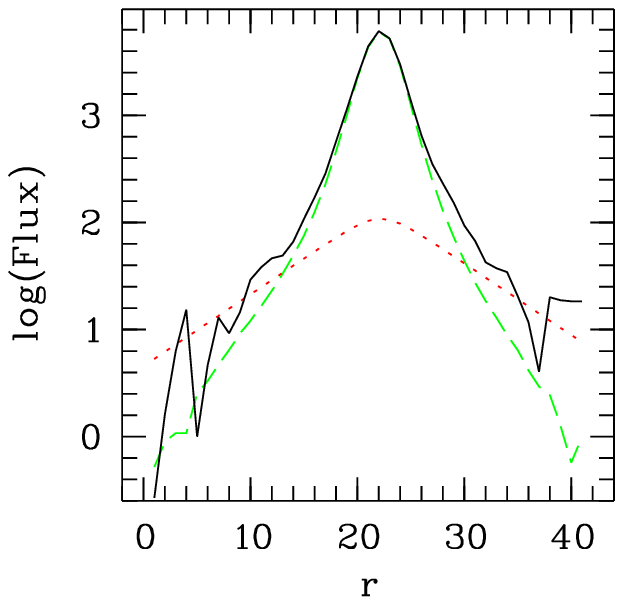}
\end{minipage}
\end{center}
\caption{\label{fig:2s_scheme} 
Schematic view of modelling: 2d-spectrum (left) with spatial direction
horizontally, and dispersion direction vertically. For each row the
composite spatial surface brightness distribution is modelled (top
right). Bottom right shows the logarithm of the total surface
brightness distribution from above (solid, black) with best fitting
models of nucleus (dashed, green) and host (dotted, red).
}
\end{figure*}

The spatial component of long slit spectra is often collapsed after reduction
(for quasar hosts e.g.\ by \citealt{hugh00} and \citealt{nola01}) to extract
the full object flux. For extended objects like galaxies, however, the spatial
direction can be used to study the galaxy spatially resolved, in our case
along a slice through the centre of the QSO. Thus instead of collapsing the
spatial dimension after data reduction, we use the 2d-frames with dispersion
and spatial direction during the complete process of separation.

Our general approach to separate active nucleus and host galaxy is very
similar to the imaging case described in \citet{jahn04a}: For each wavelength
element $\lambda$ (e.g., a row of the 2d spectrum in
Fig.~\ref{fig:2s_scheme}), the spatial 1d surface brightness distribution of
the QSO consists of a superposition of point source like nucleus and extended
host. The shape of the point source is defined by the shape of the PSF at the
time of observations and at the position of the QSO on the chip at the
specific wavelength.

In our approach we first model the \textit{spatial} shape of the PSF and then
build models for the combined light distribution of both the nucleus and the
host, asuming a certain analytic shape for the host. A fitting procedure
determines the best set of parameters for these models. Due to the low S/N per
individual wavelength element, a simultaneous fit of all parameters (e.g.\
centres, shape parameter, fluxes, etc.) will not be well constrained. We thus
exploit external input knowledge on the host to constrain as many parameters
as possible, thus reducing the number of free parameters in each fit. We also
exploit that some of these parameters vary only slowly with $\lambda$, to
reduce noise. After a parameter has been fitted, it is fixed in the following
modelling steps until in the final step only the central surface brightnesses
of nucleus and host are free parameters. In this way the fitting routine makes
optimal use of the signal, the introduction of additional noise is minimised
and the fits can succeed.
\medskip

\noindent
The 2d decomposition consists of the following steps, which in the
following we will describe in detail:

\begin{itemize}
\item Spatial definition of the PSF with the fit of an analytical
funtion to each row of a PSF star. This includes centering,
determination of a Moffat shape parameter $\beta$ and width $\alpha$
for all $\lambda$. $\longrightarrow\mathrm{PSF}(x,\lambda)$
\item Creation of an analytical look-up-table correction, to account for
intrinsic differences in the shapes of PSF model and the ``true'' PSF, using
the residual of the PSF modelling. $\longrightarrow\mathrm{LUT}(x,\lambda)$
\item Simultaneous spatial modelling of nucleus, represented by the
determined PSF, and the host, represented by a exponential disk or
de~Vaucouleurs spheroidal. 
\begin{itemize}
\item First the center is determined, and its variation with $\lambda$
fitted. $\longrightarrow x_0(\lambda)$
\item Morphological type and width of the host are determined
externally from broad band images. $\longrightarrow$ type and $r_{1/2}$
\item In a final step the fluxes of nuclear and host model are fitted
with all other parameters fixed.
\end{itemize}
\item If possible, emission lines are modelled separately to avoid
numerical flux transfer from nucleus to host model. 
\end{itemize}

\section{PSF definition}\label{sec:spec2dpsf}
The first step of the modelling is estimating the PSF. With current
instruments it is not possible to satisfy all three requirements: to know the
PSF at the position of the QSO on the chip, at the wavelength in question, and
at the time of observation. However, with everchanging ambient conditions the
temporal variations are large, and due to the dependence of the PSF on
wavelength, the wavelength condition is also crucial. In comparison the
variation of the PSF within the FOV is small for current multi-purpose
instruments. Thus the best results will be reached when defining the PSF at
the same time and wavelength regime as the QSO observation. This is realised
by simultaneously observing a PSF star with the QSO, a requirement that can be
realised with longslit or multi-object spectroscopy, and many targets.

If the spatial variations in the end are {\em sufficiently} small depends not
only on the instrument and its specific optical distortion pattern, but also
on the distance of the PSF star from the main target, the contrast of nucleus
and host galaxy, and the quality of externally input information about host
galaxy shape. Absolute statements are difficult to make: For faint AGN in
bright galaxies and small seeing the PSF will more often be sufficiently well
described than for compact systems and large seeing. With PSF variations of a
few percent, host galaxy light of less than a few percent of the total flux
will be out of reach. The brighter and spatially closer the PSF star is to the
target quasar, the better. We have a number of targets in
section~\ref{sec:applic} and \ref{sec:results} for which the combination of
conditions is not fulfilled and we can not separate successfully. While
depending on the exact science goal, we see no principle obstacle to use any
modern spectrograph for this type of science.

We describe the PSF as a combination of an analytical function and an
empirical correction, and do not use the empirical PSF star directly, for
three reasons: Firstly, the analytic approach allows to recenter the PSF to
different (sub-pixel) positions without any broadening. Secondly, the
convolution on a fine grid is substantially faster and more straightforward
than with an empirical PSF, because the empirical correction -- which is small
and has small gradients -- can simply be added in the end. Last but not least,
while we did not yet implement this, the analytical approach in principle
allows to model spatial PSF variations if more than one PSF star is observed
with the target.

The PSF modelling process schematically consists of:
\begin{itemize}
\item centering
\item global functional form determination
\item width determination
\item creation of an empirical look-up-table correction
\end{itemize}

\subsection{Parametric description}
The PSF is characterised by using a parametrised functional form, fitted for
each $\lambda$. Different functional forms could be used. A single Gaussian
has only one free shape parameter (width), but is too simple to describe real
world PSFs. We opted to use a Moffat profile \citep{moff69}, which showed to
match well in earlier imaging studies we performed with focal reducer
instruments \citep{kuhl04}. The expected and measured differences between the
Moffat functional shape and the observed PSFs are small and are taken care of
by the empirical correction (section~\ref{sec:spec2dlut}). Other functional
descriptions could have been used, e.g.\ two Gaussians, with then somewhat
different empirical look-up-table corrections, but without substantial
differences in the resulting PSF.

In the Moffat profile
\[
\mathrm{PSF}(x) = I_0 \left[1+\left(\frac{x-x_0}{\alpha'}\right)^2\right]^{-\beta}.
\] 
$x_0$ is the centre of the function, the parameter $\beta$ describes
the ``winginess'' of the profile, the relation of flux contained in
wings compared to the nucleus. $\alpha'$ is a measure of the width of
the Moffat. We use a modified description
\[
\mathrm{PSF}(x) = I_0 
\left[1+\left(\frac{x-x_0}{\alpha}\right)^2 (2^{1/\beta}-1)\right]^{-\beta}
\]
that makes $\mathrm{FWHM}=2\alpha$.

For $\beta\rightarrow \infty$ the Moffat function reduces to a
Gaussian. Atmospheric turbulence theory predicts a shape for the seeing that
can be fitted with $\beta=4.765$ \citep{sagl93,truj01}, although real data
show larger wings (lower $\beta$) due to telescope imperfections. Tests for
our image modelling showed a good fit of the shape to the data in most
cases. We find a range of $1.8<\beta<3.5$ for the spectroscopic data.

\subsection{Centroid $x_0$}
The centroid of an observed object on the detector chip will vary with
wavelength, when the object is not observed in the paralactic angle. Because
of the constraints on rotator angle imposed by the orientation of the long
slit or slitlet unit this is generally not possible. Thus first of all the
centroid is determined for each $\lambda$. For this task a Moffat function is
fitted at each wavelength, though any symmetric function showing a maximum in
the centre could be used (see below). We use a Moffat function with $\beta$
and width $\alpha$ set to initial values, and allow the centroid $x_0$ and the
central flux as free parameters.

\begin{figure}
\includegraphics[bb = 55 510 345 771,clip,angle=0,width=\colwidth]{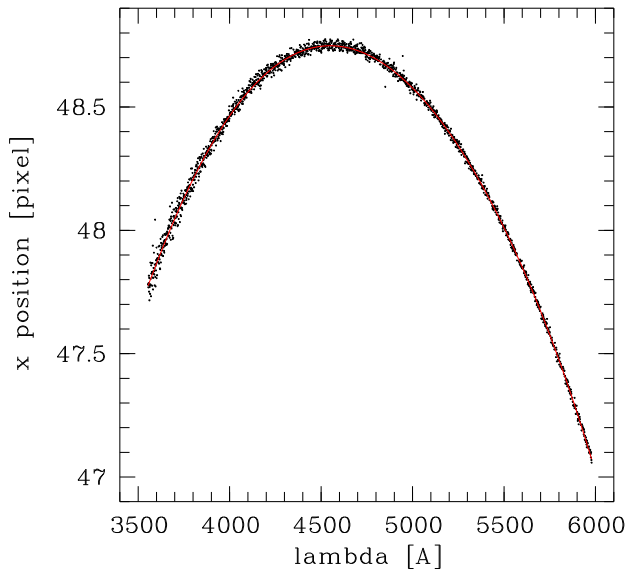}
\caption{\label{fig:centers} 
Example of PSF star centering: Plotted are the individually determined centres
for each $\lambda$ (dots). The solid line is a polynomial of degree 7 fitted
to the points. This $x_0(\lambda)$ is used as the best estimate for the centre
in the following modelling steps.
}
\end{figure}

Finding the best centroid is done in two iterations. First we use a numerical
Levenberg-Marquard least square algorithm \citep{pres95} with no subsampling
of the initial pixels, to get a robust initial estimate for $x_0$. After
repeating this for all $\lambda$ we use the fact that $x_0$ is varying only
very slowly with wavelength. Thus changes between adjacent rows will be of the
order of 1/100th of a pixel in a high resolution 2d spectrum. We use this fact
by fitting a polynomial of typical degree 3--7 to the raw $x_0(\lambda)$ to
strongly reduce the noise (Fig.~\ref{fig:centers}).

The second iteration uses these centroids as input for a downhill simplex
algorithm \citep{pres95} that again determines a centroid value. As for the
imaging case described in \citet{jahn04a} we now use a division into a finer
subpixel grid to compute the PSF model to avoid artifacts from defining the
function only on the comparably coarse pixel grid. The number of subpixels is
in general only limited by computing power. For the strong gradients in case
of a narrow PSF, it is advisable to use at least 50 subpixels. Beyond 50
subpixels the improvements become negligible.

After this step again a polynomial is fitted for $x_0(\lambda)$, using the
Levenberg-Marquard fit result as a zero-line for clipping outliers. This
relation is now being used as the best estimate for $x_0(\lambda)$. In the
next steps $x_0(\lambda)$ is fixed to this relation and is no longer a free
parameter. In this way centering precisions of the order of 1/100 of a pixel
are possible.

As mentioned above, for the inital centering it is not important to know the
shape parameter $\beta$ and the width $\alpha$ precisely. As the PSF can be
assumed to be symmetrical, the centering will yield good results, as soon as
the initial width estimate is in the right ballpark. For $\beta$ an initial
value of 2.0--2.5 is usually appropriate. The initial guesses for width and
central flux as well as the degree of the smoothing polynomials have to be
chosen individually for each spectrum.

\subsection{Shape parameter $\beta$}
The next step after centering is to estimate a global value for the parameter
$\beta$ of the Moffat function. Within limits the general shape of the PSF,
i.e.\ the ``winginess'' of the Moffat function, should not change with
wavelength. In any case the PSF shape is not perfectly described with a simple
two-parameter function, thus residuals are to be expected (they are attempted
to be removed with an analytical look-up-table, described later in
section~\ref{sec:spec2dlut}). Tests show that $\alpha$ and $\beta$ are
directly correlated when fitting a given shape (Fig.~\ref{fig:psfwidthfit}),
thus a $\beta$ changing slightly with $\lambda$ can be compensated by a
different width $\alpha$, without dramatically worsening the quality of the
fit.

With fixed $x_0(\lambda)$ we compute global reduced $\chi^2$-values, i.e.\
cumulated for all $\lambda$, for different values of $\beta$, with $\alpha$
and central flux as free parameters in a downhill simplex fit. A bracketing
algorithm is used to find the global $\beta$ with the lowest resulting
$\chi^2$ and subsequently $\beta$ is held fixed. If we assume that the PSF
were constant over the FOV, $\beta$ is then valid for PSF \emph{and} QSO as it
describes the functional form of any point source at the time of observation.

However $\beta$ might change for different objects, i.e.\ time of
observation, and also different grisms. We did not find a systematic
dependency on central grism wavelength, thus we conclude that the
changing ambient atmospheric conditions dominate the value of $\beta$
if certain conditions are met. We discuss these conditions and the
influence of the PSF star brightness below in section~\ref{sec:spec2dsn}.

\subsection{Width $\alpha$}
\begin{figure}
\includegraphics[bb = 64 510 342 774,clip,angle=0,width=\colwidth]{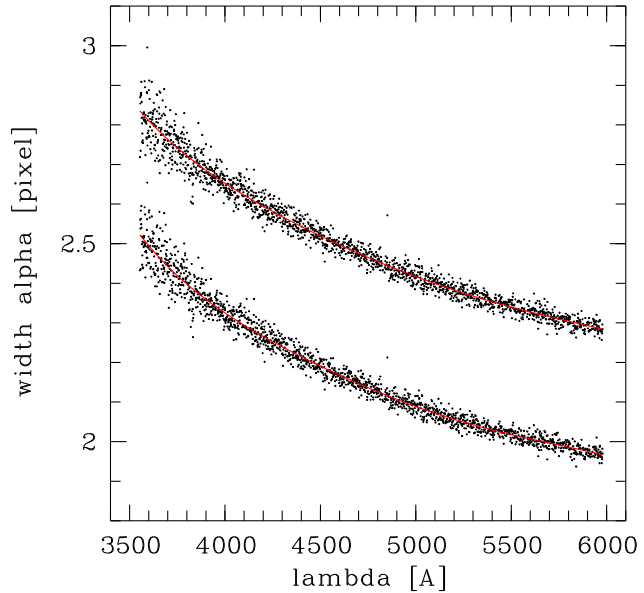}
\caption{\label{fig:psfwidthfit} 
Example of fitting the width $\alpha$ for a PSF star: Dots are again
the individually determined $\alpha$ values for each $\lambda$. The
solid line is a polynomial fit. This $\alpha(\lambda)$ is used as the
best estimate for the width of the PSF in the QSO modelling steps. As
an illustration for the correlation of $\alpha$ and $\beta$ the fits
for two values of $\beta$ are given. Top: $\beta=3.5$, bottom:
$\beta=1.75$. There is a systematic offset of $\Delta\alpha\sim0.3$,
changing slowly with $\lambda$.
}
\end{figure}

As for the centering, the width $\alpha(\lambda)$ is determined in two
steps. Centroids $x_0(\lambda)$ and $\beta$ are now fixed to their
best estimates, and only $\alpha$ and central flux remain as free
parameters. Using global initial guesses, first a robust
Levenberg-Marquard fit without subsampling is used for each $\lambda$,
followed by a polynomial smoothing fit. Then this relation is input as
an initial guess for a fully subsampled downhill simplex fit, again
followed by a polynomial fit over $\lambda$ with clipping of outliers.
The width is changing faster with $\lambda$ than the centroids (see
Fig.~\ref{fig:psfwidthfit}), over the whole spectroscopic range
changes of 30\% were observed for some objects. The change seems to be
mainly due to a change in shape due to the different atmospheric
refraction, depending on wavelength.

\section{PSF variations}\label{sec:variations}
At this point all parameters $\beta$, $x_0(\lambda)$, and $\alpha(\lambda)$
are determined. With this stepwise approach the effects of noise and numerical
artifacts are minimised. However, we initially requested as a third
requirement to estimate the PSF at the position of the QSO in the FOV, not at
the position of the PSF star. In imaging, we notice variations of the PSF
parameter over the FOV, so how strong is the effect in 2d spectroscopy?

Here the problem is simpler compared to imaging: we only have one, not two
spatial coordinates. Thus the variation of ellipticity, position angle and
width of the PSF is effectively reduced to a variation of the width. Any
change in the other spatial coordinate has only an effect across the width of
the slit and thus a (negligible) influence on the spectral coordinate, and
none on the spatial.

In the usual case that very few PSF stars are observed simultaneously with the
QSO, in many cases only a single one, it is not possible to model PSF shape
variations over the field. For crowded fields this is a theoretical
posibility. The magnitude of the variation will depend on the instrument and
its optical layout. We performed a comparison for one of our objects (see end
of next section) that showed that the effects are not very large in this
case. In section~\ref{sec:efoscpsf} we will comment on focal reducer type
instruments, and their general PSF instability.

\subsection{Sensitivity to $\beta$ and S/N effects}\label{sec:spec2dsn}
Two further sources for errors in the PSF shape are the sensitivity to the
exact value of $\beta$ and dependence on the brightness or S/N of the PSF
star. Both effects are connected. With decreasing S/N the wings of the PSF
will more and more vanish in noise, thus the wings will have less weight in
the fit. We already noted the correlation of width $\alpha$ and shape
parameter $\beta$ (Fig.~\ref{fig:psfwidthfit}). When determinining the shape
of the PSF star for decreasing S/N, $\beta$ will increase and so will
$\alpha$, leading to a similar just less wingy shape. So there are two
separate effects to be investigated: Variations in shape due to a different
$\alpha$--$\beta$ combination, e.g.\ due to different S/N, and variations in
shape for a fixed $\beta$ and different S/N. The first appears when PSF star
and QSO nucleus differ strongly in brightness.  The second is valid for every
PSF star, because a global $\beta$ is determined, but the flux and thus S/N
varies with $\lambda$.

\begin{figure}
\includegraphics[bb = 51 282 395 772,clip,angle=0,width=\colwidth]{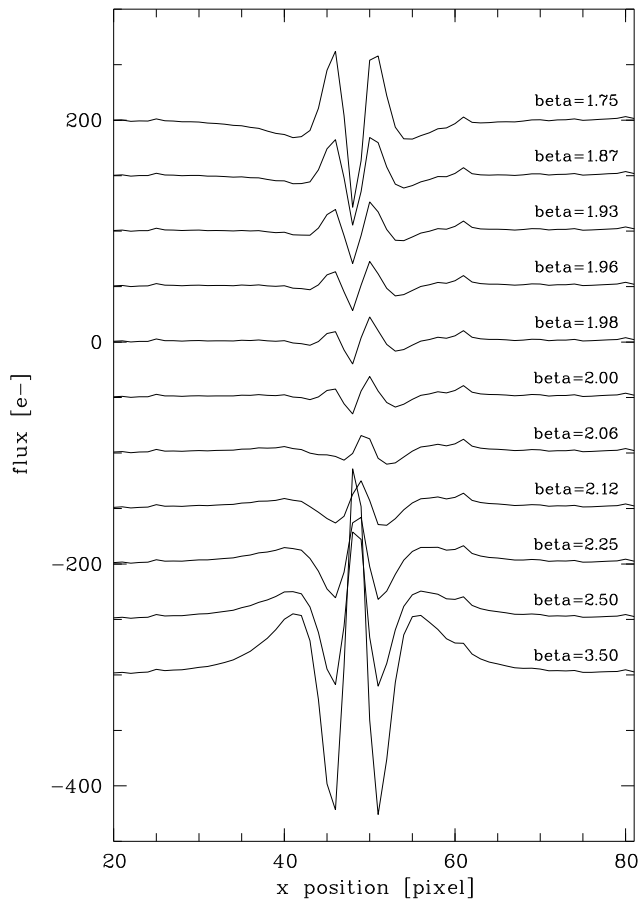}
\caption{\label{betares} 
Dependence of the PSF residual on the shape parameter $\beta$. Plotted
is the spatial residual averaged over all wavelengths for the PSF star
of HE~0956--0720 (grism 600B) for different values of $\beta$. The
curves are offset for better visibility. The magnitude of the residual
has to be compared to a central flux of the PSF star of in this case
4600 $e^-$, or 180 times more than the structures in the best fit
($\beta=1.98$). Size and structure of the residual vary from object to
object.
}
\end{figure}

An analysis of the residuals after removal of the PSF model shows the
magnitude of the effects: If for a given PSF star $\beta$ is varied by
$1.75\le\beta\le3.5$, which represents the complete range of $\beta$ values
found for our \emph{whole sample} observed with VLT FORS, the differences in
the residuals can be significant (Fig.~\ref{betares}). For any single object
the possible range in $\beta$ can be established much better. For a range of
$\Delta\beta<0.1$ the difference is of the order of one $e^-$/pixel, thus for
a ``good'' bright PSF star this is small compared to the total PSF star flux.

We also tested the change in $\alpha$ for different S/N and a fixed $\beta$,
which simulates different brightness of PSF star and nucleus.  In a frame
where two PSF stars of different brightness were available, we fixed $\beta$
and determined the width $\alpha$ for both stars. In a test case with a
difference in brighness of a factor of three, the widths are virtually
identical down to the low S/N regime of the spectrum. The only difference in
the mean relations is due to a non-gaussian scatter of the individual
$\alpha(\lambda)$ points for the fainter star, but the magnitude is not
significant. With consistent shapes down to this level, we can expect it to be
well defined for all $\lambda$ with substantial flux. Quantified as a S/N, one
should be on the safe side with a S/N in the central pixel of $>$20.

This agrees well with results for simulated PSF stars of different S/N. We
created a noiseless model PSF star with known width, added noise for different
amounts of sky background and fitted the width with the technique described
above. We used 0.15, 1.0, 7.1 and 71.0 times the typical sky background noise
for our objects. The width does change systematically, but a very good
reconstruction of the width better than 5\% can be found for S/N of the
central pixel of better than 10--20. In the modelling of the EFOSC sample,
further evidence was found that stars below a minimum S/N of $\sim$20 in the
central pixel can create artificially large $\beta$ values, largely
independent of ambient conditions.

\subsection{Look-up-table correction}\label{sec:spec2dlut}
Besides variation of the PSF with S/N and position over the FOV, there is also
a systematic mismatch between shape of the PSF star and its the two-parameter
Moffat function model. The difference between PSF and model is the same for
PSF star and QSO, assuming a constant PSF shape. We correct the majority of
this mismatch by constructing an empirical look-up-table (LUT) correction,
consisting of the modified residual of the PSF modelling process.

The residual is smoothed using a moving average along the dispersion direction,
in order to reduce noise. This can be done because we assume the residuals
forming the LUT to be strongly correlated in adjacent $\lambda$. In the same
process it is normalised with the flux of the PSF star, and rebinned to match
centre $x_0$ and $\lambda$ of the QSO.

Tests show that LUTs from different PSF stars vary in their details, but show
a generally identical shape. The resulting LUT applies corrections on the
order of a few percent, in the best cases less than 1\%. Thus the second-order
errors in LUT difference are very small.

In total we have a smoothed empirical correction function
$\mathrm{LUT}(x,\lambda)$ to complement the analytical PSF description
$\mathrm{PSF}(x,\lambda)$.


\section{Modelling nucleus and host galaxy}\label{sec:spec2dqso}
Modelling the QSO follows the same principles used for the PSF
star. Parameters are estimated, their variation with $\lambda$ fitted, and
this relation fixed in subsequent steps, reducing the number of free
parameters in each step as much as possible.

The centering is directly identical to the PSF modelling, using a single
Moffat function. In all following steps two components are modelled: a PSF
like component representing the nucleus and an extended component for the
host. The S/N per wavelength element is generally so low that simultaneously
determining parameters for a third component, e.g.\ gas disk or separate
bulge, is not possible (but we will discuss this possibility for the treatment
of emission lines in section~\ref{sec:emissionlines}). For the host we assume
a one-dimensional exponential disk or de Vaucouleurs spheroidal profile, as in
the imaging case, but in a one-dimensional form:

\begin{eqnarray}   
 \mathrm{SPH}(x) &=& \mathrm{SPH}_0
 \exp\left[-7.67\,\left(\frac{\left|x\right|}{r_{1/2}}\right)^{1/4}\right]
 \label{eqn:sph}\\
 \mathrm{DISC}(x) &=& \mathrm{DISC}_0
 \exp\left(-1.68\,\frac{\left|x\right|}{r_{1/2}}\right) \label{eqn:disc}
\end{eqnarray}   
These models also have two parameters, central flux $I_0$ and half
light radius $r_{1/2}$.

The creation of the spatial models is again done on a fine subpixel grid. For
each point on this grid the value of the PSF is computed, as the sum of
analytical Moffat function plus empirical LUT. The host model is convolved
with the PSF as a discrete convolution at these grid points. After
computation, the grid points are binned back onto the (coarser) pixel grid of
the data frame. This convolution is the main reason for using a downhill
simplex algorithm.  As it has to be done numerically, no derivatives are
available.

\subsection{Scale length of the host galaxy}
In the step after centering the half light radius and the galaxy type
(disk/spheroidal) are determined. This can either be done via fitting, with
$x_0(\lambda)$ fixed and $I_0$, $r_{1/2}$ and central flux of the nucleus as
free parameters. But due to the low S/N this is a source for failure in many
cases. The host can be outshone by the nucleus by a factor of 100 or more. In
spectroscopy by sampling just along a slice through the QSO which contains
most of the nuclear but only a small fraction of the host light, the total
nucleus-to-host flux ratio is boosted. Wherever available, we use external
information on host galaxy type and $r_{1/2}$ to overcome this problem, by
obtaining an auxilliary image for any object and determining the morphological
parameters as described in \citet{jahn04a}.

The price for a deep 2d image of a QSO host to determine morphological host
parameters is small compared to the price of the spectrum itself. Thus for
little extra observation time -- for the redshift range accessible to this
spectral modelling, $z<0.5$, one can use 2m-class telescopes, or invest
30--60~s of 8m-class time -- the results are greatly enhanced, if not only
made possible in many cases.

Thus the functional form and half light radius are set from external
sources. As we noted in \citet{jahn04a}, the scale lengths of galaxies can be
variable with $\lambda$ due to colour gradients. Because usually no multiband
imaging data is available we have to assume it to be constant. $r_{1/2}$ and
$I_0$ are strongly correlated parameters. Even though their individual values
are not well constrained for low S/N, their product in the form of the total
flux can be estimated with a much higher precision \citep{abra92,tayl96}. Thus
the quality of the fit and total host flux will not strongly depend on the
exact value of $r_{1/2}$.

If $r_{1/2}$ should after all be taken from the spectrum, one way to increase
the S/N is to coadd parts or all of the spectral dimension, with care taken
about the varying centroid, and all emission lines omitted, and model the
resulting 1d frame. In total the S/N of the host is greatly
increased. Depending on S/N, even a varying $r_{1/2}$ could be determined by
splitting the spectrum into two or more parts.  This process can be done
iteratively: Modelling is first done with an initial guess. The resulting host
spectrum is used to extract a better spatial model to be used in the second
iteration etc.

\subsection{Final step: the host spectrum uncovered}
In the final step just two parameters are free, the central fluxes of host and
nucleus, with all other parameters fixed. The resulting model of the host is
not used for further analysis, only the 2d nuclear model image is used as the
best estimate for the nuclear component and subtracted from the initial 2d QSO
spectrum. The result is a residual host galaxy frame on one hand including all
PSF removal imperfections, but on the other hand containing all deviations
from symmetry like rotationally shifted lines that are not modelled by the
symmetric model, and differences in actual morphology.

This frame can now be subjected to standard 2d spectra calibration and
analysis techniques. The 1d spectrum is extracted using the Horne
optimal extraction algorithm \citep{horn86}.

\subsection{Emission line treatment}\label{sec:emissionlines}
Refinements have to be applied in the treatment of emission lines. While the
broad emission, narrow emission and continuum of the QSO have an identical
spatial shape, this is not the case for the host. The spatial distribution of
the interstellar gas, responsible for narrow emission lines visible in the
host, does not need to be identical to the distribution of stellar
emission. This is particularly obvious in elliptical galaxies, where rotating
gas disks can be embedded in the spheroidal distribution of stars.

This has the consequence that at wavelengths with a substantial contribution
of gas emission lines, the two-component model of nucleus plus host is a wrong
assumption. E.g.\ in the case of an elliptical galaxy with a gas disk, the gas
emission will spatially be more extended and less peaked than the stellar
emission. To account for the extra flux at larger radii the numerical
algorithm will assign extra flux to the more extended of the two components,
i.e.\ the host. This boosted host model will match the gas emission well at
the radii in question, but will overestimate it in the center. This will be
compensated by the nuclear model, that as a result will be underestimated.

Three solutions can be used, for different situations:

\begin{figure}
\begin{center}
\includegraphics[bb = 0 0 352 177,clip,angle=0,width=\colwidth]{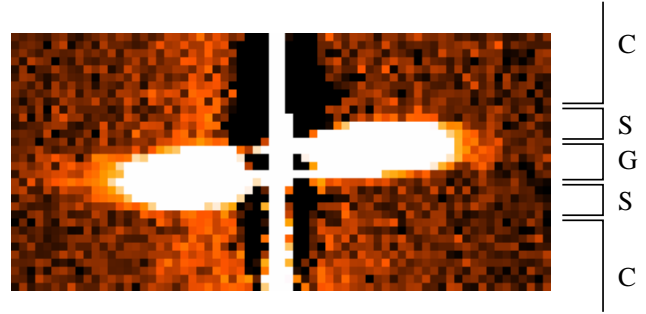}
\end{center}
\caption{\label{fig:linemodeling} 
Emission line in a resulting host spectrum, dispersion vertical,
spatial coordinate horizontal. The two sides are rotationally shifted
in wavelength due to rotation of the gas disk. Three fundamentally
different regions can be identified: Pure continuum emission (``C''),
single sided emission line (``S'') and double sided gas emission
(``G''). For all three regions different modelling conditions apply.
}
\end{figure}

\begin{itemize}
\item Single sided modelling: In the case of a rotating gas disk, both
sides of the line are rotationally shifted away from the nominal
position, one side blueward, the other redward. In the case of a
narrow line and a rotation curve that does not strongly decrease with
radius, e.g.\ the case for gas disks in elliptical galaxies, only
parts or even none of the single sided components overlap in
wavelength (marked ``G'' in Fig.~\ref{fig:linemodeling}).

In the non-overlap wavelength elements (marked ``S''in the same
figure) it is possible to model just one side of the 2d spectrum. With
fixed PSF, $x_0$ and $r_{1/2}$, the nuclear model is determined just
for the side free of line and only continuum emission. This nuclear
model is then valid also for the other, line emission ``contaminated''
side of the spectrum. In this way an emission line with
non-overlapping sides can be split up into two parts and the single
sided nuclear models can be merged with the full models for all other
$\lambda$.

\item If the two sides of the line do have some overlap, the
underestimate for the nucleus might be seen in the resulting
spectrum. It could be manifested as a absorption like feature where
none is expected, or as a disturbance in the spectrum. In this case
the oversubtraction in the nucleus model can be estimated and a better
nuclear contribution constructed. This however is difficult to judge.

\item A third approach, not yet implemented, is the use of a third model
component in the emission line. Here the S/N is large and the stellar
continu\-um can be estimated from adjacent spectral intervals (marked ``C'' in
Fig.~\ref{fig:linemodeling}). Even with nuclear contamination a good scale
length for the gas disk can be obtained from the radial shape, when excluding
the central spatial part of the line. The stellar contribution would enter
only as a constant component. With a set gas component scale length, again
only two free parameters, $I_0$ of nucleus and gas, would be present.
\end{itemize}

\subsection{Continuum flux transfer}
If two spatially different components exist in the object under investigation,
and the S/N is sufficiently high, this modelling procedure will find those
components. However, a signal of the host might be found only in the brightest
parts of the spectrum, e.g.\ a narrow gas emission line, while the stellar
continuum is outshone by the luminous nucleus. If the variation of the PSF is
large and the host compact, errors in the PSF will be compensated by the
routines with changes in the host model flux. As a result, more or less strong
flux transfer can occur from the nucleus to the host or vice versa. This can
go as far as negative fluxes for a part or all of the resulting host
spectrum. In this case the errors in the PSF determination are larger than the
positive signal of the host. When this happens, one can still extract some
information from the resulting two component spectra, because the spectrum was
still separated into a wider and a more compact component. A solution is to
add increasing fractions of the nuclear component, possibly of wavelength
dependent size, to compensate for the modelling error, and make the host flux
positive again. After such a procedure, it will not be possible to extract
detailed host continuum information, but host line ratios, rotation curves
etc.\ can still be extracted to some extent.

\subsection{Quality of fit diagnostics}\label{sec:spec2dquality}
Similar as for the PSF star modelling, the models used for the galaxy can
never describe all of the data. Spiral arms, bulges in disk galaxies or
asymmetries are not accounted for. Thus even with a LUT correction for the
nuclear component, there can be substantial residuals of data minus best
nucleus and host model. This is no particular problem as the modelling is only
used to estimate the \emph{nuclear} component. As long as the fit is ``good'',
the resulting \emph{model} of the host is irrelevant. To asses the
``goodness'' and quality of a modelling run, six fundamentally different
diagnostic principles are available:

\begin{itemize}
\item Boundary conditions: Some general boundary conditions from basic
physical principles apply, that, when not hardwired into the modelling
routine, can be used to check results. The most important are
\begin{itemize}
\item host flux should be zero or positive for all $\lambda$, within
the limits of noise,
\item forbidden lines cannot appear in absorption.
\end{itemize}
\item The residuals of the fit can be used to detect misestimates of
the host's scale length. When comparing the residuals it is possible to
discriminate between different scale length values. Too compact
models show residual flux in the wings.
\item The nuclear broad emission lines should not appear as strong
mirrored absorption lines in the resulting host flux. Broadened
absorption lines can occur in galaxy spectra, but not of a width of
several 1000~km/s as the permitted nuclear emission lines.
\item Broad band colours existing from imaging studies must be
reconstructable within the errors of calibration.
\item If more than one PSF star of suitable S/N exists, the results
from modelling with both stars can be compared.  
\item Finally, if several images exist for the same object, taken with
different grisms, the fit results in overlapping spectral regions can
be compared agains each other.  
\end{itemize}

\noindent
It has to be kept in mind at all times that every algorithm, also the
one described, is limited by S/N. It can not extract information that
simply is not there.

\section{Spectroscopic decomposition: example objects}\label{sec:applic}
Applying this method to actual data is the crucial test. We want to show the
detailed results of one example from each dataset and summarize more of the
science results after that.

All objects are low-$z$ quasars, $z<0.2$ and $z<0.3$ respectively for the
EFOSC and FORS1 instruments with apparent total magnitudes from
$V\sim13.5\ldots16.5$ (see Tab.~\ref{tab:sampleprops}). Integration times
ranged from 20~min with FORS to typically 30--60~min with EFOSC, split up into
several exposures. Slit sizes, spectral and spatial resolution were
2''/$\sim$250/0\farcs314 and 1''/$\sim$700/0\farcs2 for EFOSC and FORS
respectively. For FORS three grisms were used to cover the range from
4000--8000~\AA.

The MOS multi-object mode used with FORS was effectively used as a more
flexible long-slit thus the data is not fundamentally different for those two
data sets. In the observations the slit angle and mask layout were set so that
at least one bright accompanying PSF star was observed simultaneously with the
QSO in the longslit. The PSF star was chosen to be close to the object,
bright, but not too bright to reach the non-linear regime of the detector or
even saturate. The airmass for all observations was at or below 1.1.

The EFOSC sample is a subsample of the objects investigated with broad band
imaging in \citet{jahn04a}. Thus apart from the astrophysical analysis this
sample can be used to crosscheck the output of our modelling with results from
broadband imaging. For the FORS objects only in the $H$ band decent data was
available, the imaging observations are part of a different project
\citep{kuhl03}. We chose the objects HE\,0952--1552 ($z=0.108$,
$V_\mathrm{tot}=15.8$, nucleus/host$(V)=1.2$, nucleus/host$(H)=0.6$) as an
example for the EFOSC data quality and HE\,1503+0228 ($z=0.135$,
$B_\mathrm{tot}=16.3$, $H_\mathrm{tot}=13.6$, nucleus/host$(H)=1.5$) for
FORS. Both objects are disk dominated.

\subsection{PSF modelling}\label{sec:efoscpsf}
EFOSC and FORS are focal reducer type instruments. As is known the complex
optical layout with multiple optical elements in the light path will produce
geometric distortions in the focal plane and a shape of the PSF depending on
position in the FOV. We investigated this effect in detail in \citet{kuhl04}
for the imaging case, and in principle the same problem applies to the
spectroscopy case. Here, however, we do not have the possibility to model
these distortions by analysis of a large number of PSF stars -- generally only
one or two stars are observed. It can only be attempted to minimize the
effects by observation design.

Generally, the difference in shape between two points in the FOV will increase
with increasing distance. Therefore it is advisable to observe PSF stars as
close as possible to the QSO. However, as noted in section~\ref{sec:spec2dsn},
we require a minimum brightness for the stars, to reduce the uncertainty in
PSF shape determination. Thus for most host galaxy studies there is only a
small choice of stars to be used. For these samples it was attempted to find a
good compromise between brightness of the star and the distance to the
QSO. Generally instruments should be preferred that show a stable shape of the
PSF. This can even be of higher priority than light collection power. Too
strong variations can not be compensated by higher S/N. For our samples we
found that the distortions in FORS are much small than for EFOSC, as expected
for the younger and higher quality optics.

\subsection{Decomposition results}\label{sec:efoscmodelling}
Modelling was carried out as described in the previous sections. We used 80
subpixels per input data pixel in the spatial coordinate, and the LUTs were
created by smoothing over a radius of 30 pixels in wavelength.

\begin{figure*}
\begin{center}
\includegraphics[bb = 53 75 284 772,clip,angle=-90,width=18cm]{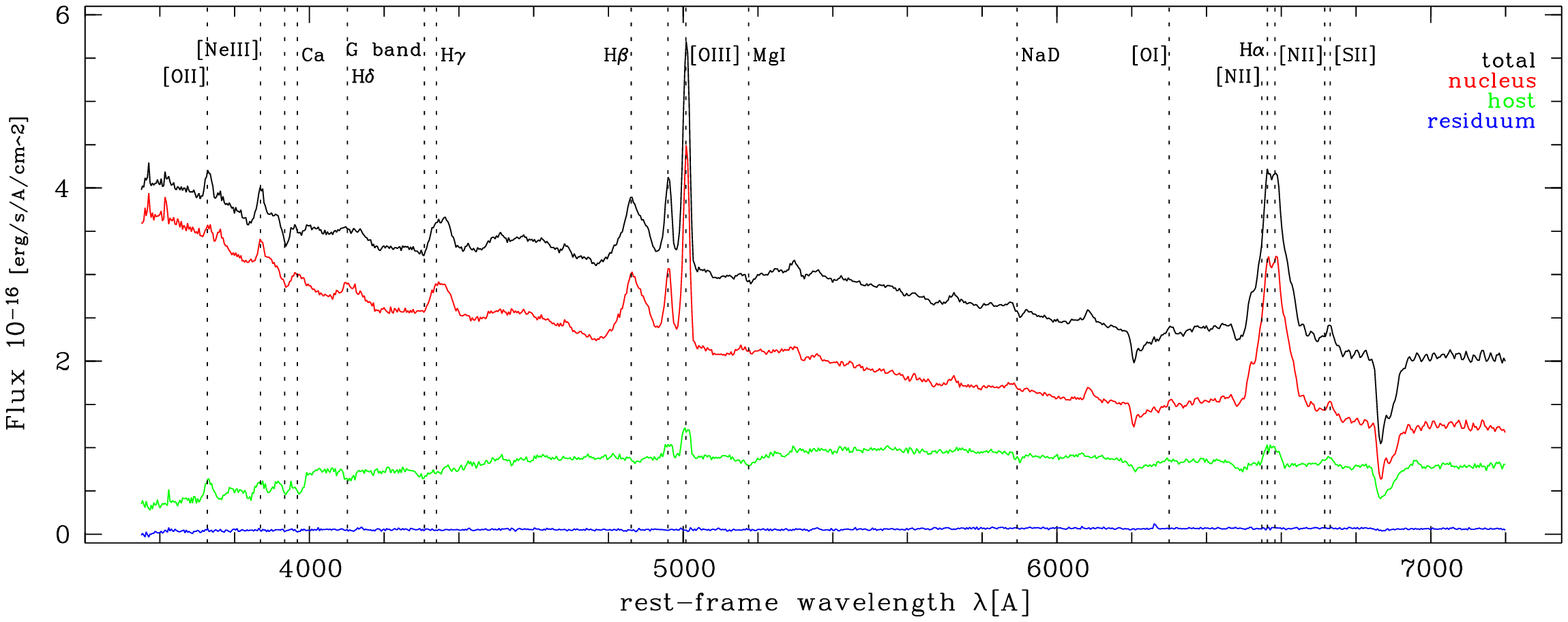}
\includegraphics[bb = 53 75 315 772,clip,angle=-90,width=18cm]{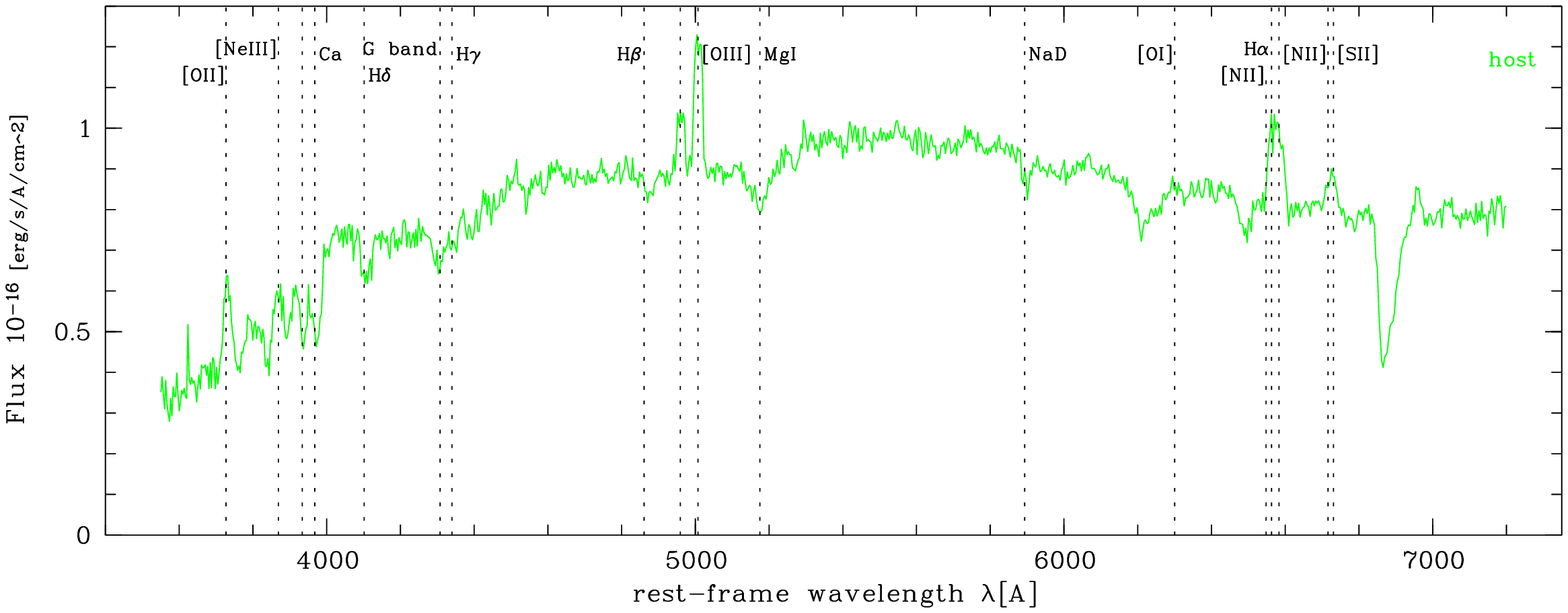}
\end{center}
\caption{\label{fig:efosc0952_modelling} 
Modelling result EFOSC sample: HE\,0952--1552. Shown are, in the top panel,
extracted spectrum of the total QSO (top black line), nucleus (red line), host
after subtraction of the nuclear model (green line) and residual after
subtraction of nucleus and host model (blue). The bottom panel shows the host
spectrum alone. Positions of major emission/absorption lines are marked. Given
is the rest-frame wavelength. For presentation the spectra have been slightly
smoothed to enhance the visibility of spectral features.
}
\end{figure*}

\begin{figure*}
\begin{center}
\includegraphics[bb = 54 109 198 571,clip,angle=-90,width=16cm]{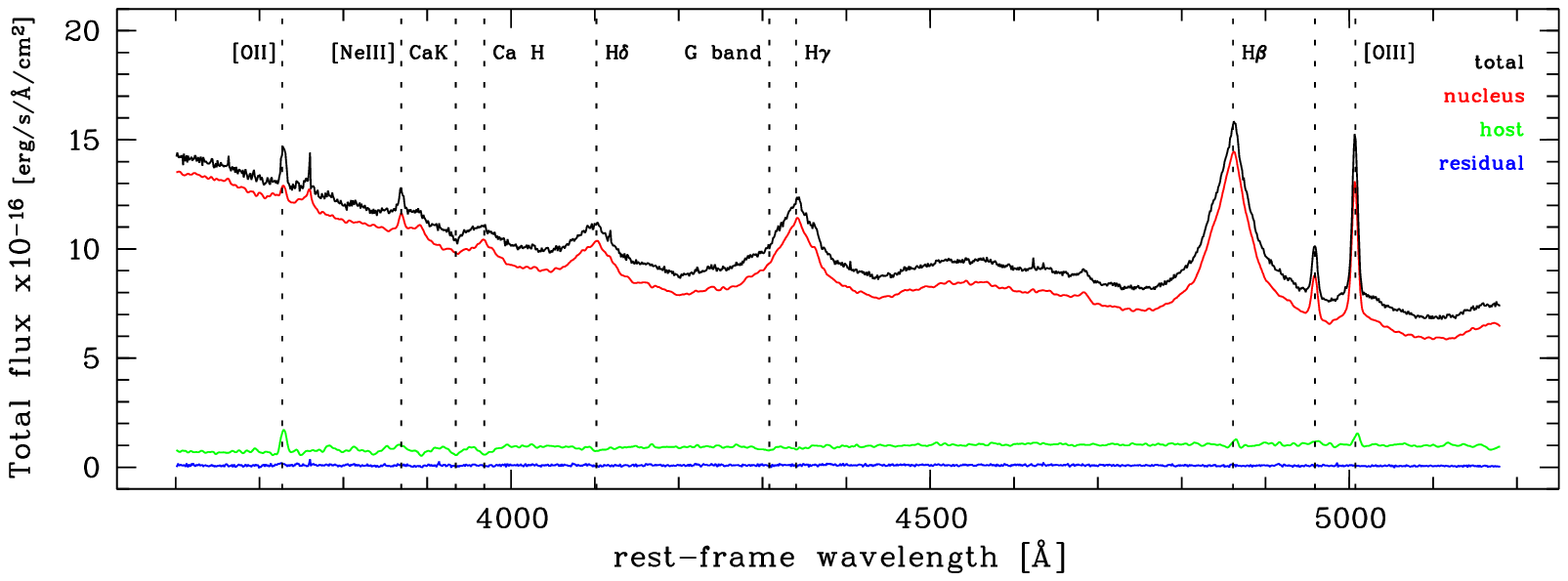}
\includegraphics[bb = 46 109 222 571,clip,angle=-90,width=16cm]{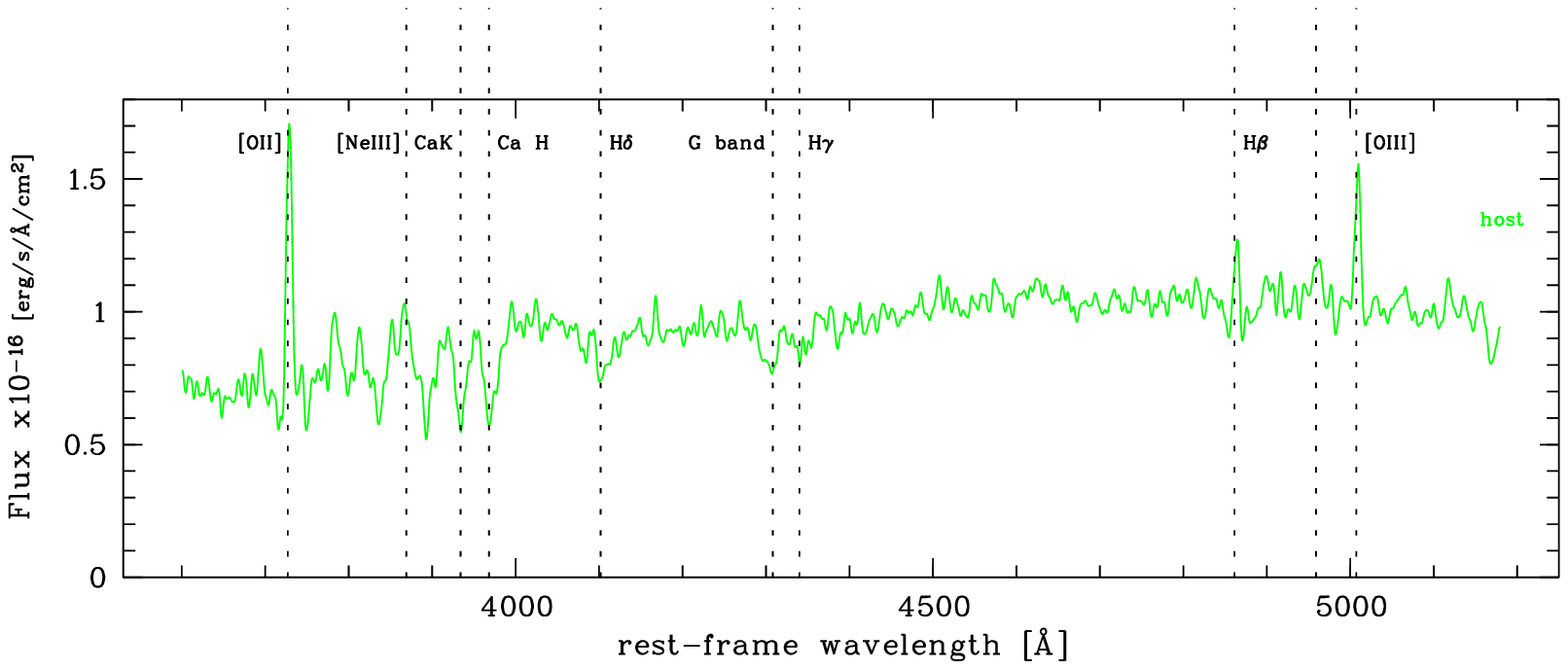}
\end{center}
\caption{\label{fig:1503_modelresult} 
Modelling result FORS sample: HE\,1503+0228. Lines as in
Figure~\ref{fig:efosc0952_modelling}.
}
\end{figure*}

Figures~\ref{fig:efosc0952_modelling} and \ref{fig:1503_modelresult} display
the modelling results for each of the two example object, in the form of the
extracted 1d spectra. The top panels show (from top to bottom) the original
spectrum of the QSO (black), the model for the nucleus (red), the resulting
host spectrum after subtraction of the nuclear model (green) and the residual
after subtraction of nuclear and host model (blue). The bottom panels show a
zoom of the host spectrum.

For both objects the modelling was straight forward without further manual
intervention, i.e.\ the errors in PSF and host morphology estimation were
small enough for the S/N, contrast and spatial resolution given. There are no
obvious oversubtractions visible, i.e.\ flux transfer to the nucleus. The host
spectra are void of nuclear broad lines, while the narrow line components in
\ha and \hb are clearly present. For HE\,1503+0228 even some Balmer absorption
can be detected underneath the narrow \hb line.

\subsection{EFOSC example: HE\,0952--1552}\label{sec:0952}
In the resulting host galaxy spectrum of HE\,0952--1552
(Fig.~\ref{fig:efosc0952_modelling}) several stellar absorption lines are
visible, Ca\,H and K, G band, Mg\,I, Na\,D, and Balmer absorption in \hb to
H$\delta$. The 4000~\AA\ break is prominent. Emission lines from the ISM are
also present: [O\,II] 3727~\AA, the [O\,III] doublet 4959/5007~\AA, [N\,II]
6548/6583~\AA\ and narrow \ha also in emission, although the three components
are not clearly resolved. The [S\,II] 6716/6731~\AA\ doublet is also visible
and it, too, is blended.

\begin{figure}
\begin{center}
\includegraphics[bb=0 4 331 280,clip,width=\columnwidth]{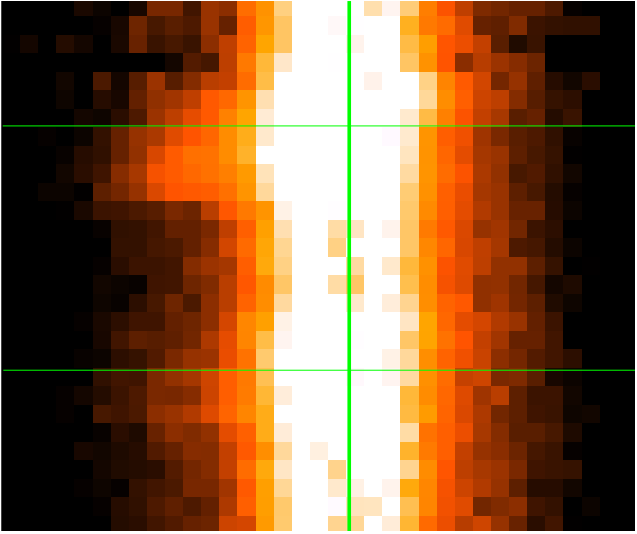}
\end{center}
\caption{\label{fig:he0952_oiii} 
HE\,0952--1552: Region of the \oiii emission line doublet at 4959/5007~\AA,
wavelength increasing from bottom top. The vertical and horizontal lines mark
the spatial and spectral centers of the lines. Visible is the effect of
rotation and a prominent asymmetric component on the approaching sides of the
two lines.
}
\end{figure}

The 2d residuals show significant structure on a level of 10\% of the host
count rate, or $<2\%$ of the count rate of the input QSO spectrum. In the
\oiii 5007~\AA\ component there are clear systematic velocity differences
visible between the two sides of the galaxy. A velocity difference between the
two sides of the galaxy along the slit can be estimated, though the \oiii
emission seems to be spatially asymmetric (Fig.~\ref{fig:he0952_oiii}).
Averaging between 1\farcs5 and 3\farcs5 radius the two components yields a
line of sight velocity difference of $\sim$350~km/s, not deprojected for slit
angle orientation and inclination.

Rotation in the absorption lines is also visible. Ca\,K shows a shift of
5.2~\AA\ between 1\farcs5 and 3\farcs5 radius of the left and right side,
corresponding to an observed rotation velocity of 200~km/s, substantially
smaller than the \oiii velocities. In conjunction with the asymmetry of the
\oiii line this poses the possibility that one side of the galaxy shows an
outflow superposed of ionized gas on the rotation pattern defined by the
stars.

The inclination is difficult to assess because of the supposed presence of a
significant bulge. From its axial ratio we estimate an inclination of
$\sim$50$^\circ$, which would be a lower limit in the presence of a bulge. In
general, quasar host galaxies rarely show inclinations above 60$^\circ$ due to
the obscuration of the central source beyond that \citep{mcle95b,tayl96}.

For this range of values, the deprojection factor for the velocity lies
between 1.15 and 1.75, depending on the angle between slit and rotational axis
of the galaxy (estimated to 0--30$^\circ$), the latter dominating the
uncertainty in this case. This leads to a deprojected velocity difference of
230--350~km/s; the best guess is $\sim$280~km/s. This value is consistent with
the rotational velocity expected for a massive bulged spiral galaxy as
HE\,0952--1552.

The rotation seen in the emission lines clearly identifies them as gas
emission and not artefacts of the separation. It is more difficult to judge
whether the strength of the emission lines determined is correct, or if
numerical flux transfer has happened from the nucleus to the host. Due to the
low resolution it is not possible to apply a single sided modelling to the
lines in this case. The model for the \oii line shows a somewhat disturbed
spectral geometry, with a small dip in the center, while the (much stronger)
\oiii lines appear completely undisturbed. Close to the nominal position of
the latter appears absorption, compatible with the Balmer absorption in \hg
and H$\delta$. But it is shifted by 12~\AA, which is not due to errors in
calibration. \ha and the \nii doublet are blended. In the host spectrum a dip
is visible at the nominal \ha position. This is also compatible with
superposed gas emission and stellar absorption.

\subsection{FORS example: HE\,1503+0228}
HE\,1503+0228 is a QSO at redshift $z=0.1350$, with a prominent disk-dominated
host with a scale length of 1\farcs9. It shows a continuum nucleus-to-host
flux ratio of 7--10 integrated over the 1\arcsec slit. With these properties
it is among the object in the FORS sample that can be studied the easiest.
QSOs with more compact host galaxies will be more difficult to separate with
larger uncertainties or requiring a better knowledge of the host's morphology.

Several prominent gas lines are visible in the resulting host galaxy spectrum,
in Figure~\ref{fig:1503_modelresult}. In the $B$ grism [O\,II], [O\,III],
H$\beta$, Ca\,H and K and Fe\,I 5270 are separately modelled with single sided
fits. A slight offset between absolute fluxes exists in the overlap regions of
$B$ and $R$ grism, of the order of 10\% at 5100~\AA.

The detected gas emission shows clear signs of rotation, so do the stellar
absorption features. The Ca\,H and K, G band absorption lines are prominent in
the $B$ grism.

\begin{figure}
\begin{center}
\includegraphics[angle=-90,width=7.5cm]{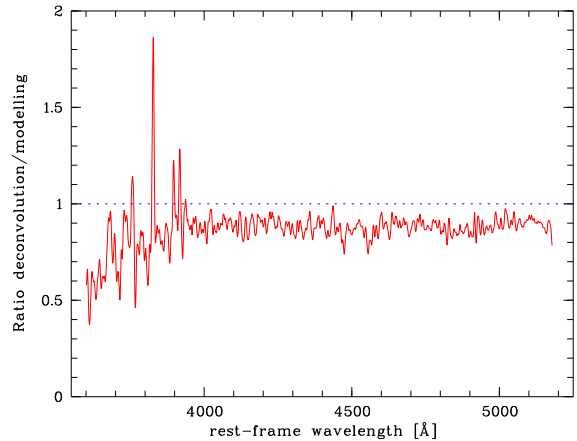}
\end{center}
\caption{\label{fig:deconv_decomp_comparison} 
Ratio of derived spectra for the host of HE\,1503+0228 from deconvolution and
th modelling decomposition described here. The two methods are consistent to
about 10\% over the higher S/N part of the data. Below 4000\AA (rest-frame)
the S/N drops and some spikes in difference appear. These are {\em not}
coincident with emission or absorption lines.
}
\end{figure}

The full FORS dataset (20 objects) and among it HE\,1503+0228 have separately
been analysed using the deconvolution method by \citet{cour00}. Results have
been published in \citet{cour02b} and \citet{leta06}. In
Figure~\ref{fig:deconv_decomp_comparison} we show the ratio of extracted host
galaxy spectra from deconvolution and modelling decomposition in the $B$ grism
for HE\,1503+0228. The two methods are consistent within about 10\% over the
higher S/N part of the data above 4000\AA. Below 4000\AA\ (rest-frame) the S/N
drops and some spikes in difference appear. However, these spikes are {\em
not} coincident with emission or absorption lines. E.g.\ at 5007\AA\ no
particular difference exists, showing that emission lines are treated
consistently between the two methods. The difference of 10\% can have
different sources. One of them is a sensitivity of the described modelling
decomposition to the input host galaxy model. If this is not constrained well
enough, a general flux transfer can occur. We have made tests with a
$\sim$30\% larger host galaxy size, resulting in 20\% higher flux values for
the host spectrum.

We can also compare line properties, e.g.\ derived line shifts. When coadding
all accessible emission lines we can compute a rotation curve, with a measured
maximum velocity of $\sim$160\,km/s. \citet{cour02b} find very similar values.
While deconvolution by \citeauthor{cour02b} seems to be able to trace the full
increase of velocities outside of a few 100~kpc (the innermost pixel), we only
resolve the velocity field starting at a radius of $\sim$2~pixels and miss the
rise.

Derived line properties are always very consistent between the two methods
for all objects in the overlapping sample (see also
Table~\ref{tab:efosc_rotation}). However, derived absolute fluxes of the
host galaxy can differ. For HE\,1503+0228 this is below 20\% and
consistent for all wavelengths, this can be stronger in other cases. A
detailed quality of fit analysis (Sec.~\ref{sec:spec2dquality}) is in each
case a prerequisite to judge if a separation was successful or failed.

\section{Science results}\label{sec:results}
So far we described the decomposition method and did quality checks. We now
also want to summarize science results on the two samples. For the FORS sample
we will restrict the description to avoid duplicating results that has already
been reported in more detail in \citet{leta04,leta06} or \citet{cour02b}.  The
results from our two methods are sufficiently similar -- whereever we
performed the same analysis we agree within our determined error bars. For
this reason the emphasis here will lie on i) stellar population fits to the
host galaxy continuum, and its comparison to broad band photometry, ii) on
peculiar velocities for the EFOSC sample, and iii) on spatial ionized ISM
distribution for the FORS sample.

As mentioned, in Table~\ref{tab:sampleprops} a summary of object properties of
the samples is shown, with redshifts, magnitudes, nucleus to host ratio,
morphological type and scale length.

\begin{table}
\caption{\label{tab:sampleprops}Objects in the EFOSC (top) and FORS samples
(bottom). Listed are redshift $z$, apparent optical magnitude (extinction
corrected; $V$ for EFOSC, $B$ for FORS), and in the near-infrared, $H$,
nucleus-to-host flux ratio n/h($H$), morphological type of the host
((E)lliptical or (D)isk dominated; ED describes objects fit better with an
spheroidal model, but a significant disk contribution can not be ruled out, DE
vice versa), and scale length $r_{1/2}$ used in modelling. HE\,1201--2409 has
a very compact early type host galaxy. It was modelled also with a disk, which
allows a better convergence of the fit. In this way at least some of the
external spatial information is used.
Except for HE\,1029--1401 and HE\,1434--1600, which are part of the
multicolour sample investigated in \citet{jahn04a}, the morphological
properties are results from \citet{kuhl03} or from Wisotzki et al., in prep.
}
\begin{center}
\begin{tabular}{llccccc}
Object &
\multicolumn{1}{c}{$z$}&\multicolumn{1}{c}{$B$,$V$}&\multicolumn{1}{c}{$H$}&\multicolumn{1}{c}{n/h ($H$)}&\multicolumn{1}{c}{Type}
&\multicolumn{1}{c}{$r_{1/2}$ [$''$]}\\
\hline
HE\,0952--1552&	0.108& 	15.8&13.4&0.6&D&2.55\\  
HE\,1029--1401& 0.085& 	13.7&12.2&2.1&E&1.95\\
HE\,1201--2409& 0.137& 	16.3&14.2&0.7&E&0.47\\
              &      &      &&   &D&3.1 \\
HE\,1239--2426& 0.082& 	15.6&12.8&0.9&D&4.4 \\
HE\,1310--1051&	0.034& 	14.9&12.7&1.2&D&3.7 \\
HE\,1315--1028& 0.098& 	16.8&14.2&0.9&D&3.3 \\
HE\,1335--0847& 0.080& 	16.3&14.1&0.45&E&2.4\\
HE\,1416--1256&	0.129& 	16.4&14.4&1.1&E&2.1 \\
\hline
HE\,0914--0031& 0.322   &16.0 &15.0 &3.5 &D &2.0  \\
HE\,0956--0720& 0.3259  &16.3 &14.3 &3.0 &D &3.0  \\
HE\,1009--0702& 0.2710  &16.4 &14.3 &1.8 &ED&0.55 \\
HE\,1015--1618& 0.2475  &15.4 &13.6 &6.0 &D &2.2  \\
HE\,1029--1401& 0.0859  &13.3 &12.2 &2.1 &E &1.45 \\
HE\,1228+0131 & 0.117   &14.7 &12.9 &5.4 &DE&1.65 \\
PKS\,1302--1017& 0.2784 &14.6 &13.6 &1.7 &E &0.6  \\
HE\,1434--1600& 0.1445  &15.1 &13.6 &1.2 &E &2.2  \\
HE\,1442--1139& 0.2569  &16.8 &14.6 &0.45 &D &1.95\\
HE\,1503+0228 & 0.1350  &16.3 &13.6 &1.5 &D &2.2  \\
\hline
\end{tabular}
\end{center}
\end{table}

\subsection{Stellar populations and broadband colours}\label{sec:pops}
Stellar populations can either be investigated by analysing diagnostic
absorption line ratios, as has been done by \citet{leta06} for the FORS
sample. Or, in a more crude way, by fitting stellar population models to the
stellar continuum, including characteristic features as the Balmer break
around 4000\AA. We want to do the latter, to extract stellar age information
in comparison to \citet{leta06}.

\begin{figure*}
\begin{center}
\includegraphics[bb = 53 103 197 571,clip,angle=-90,width=13.95cm]{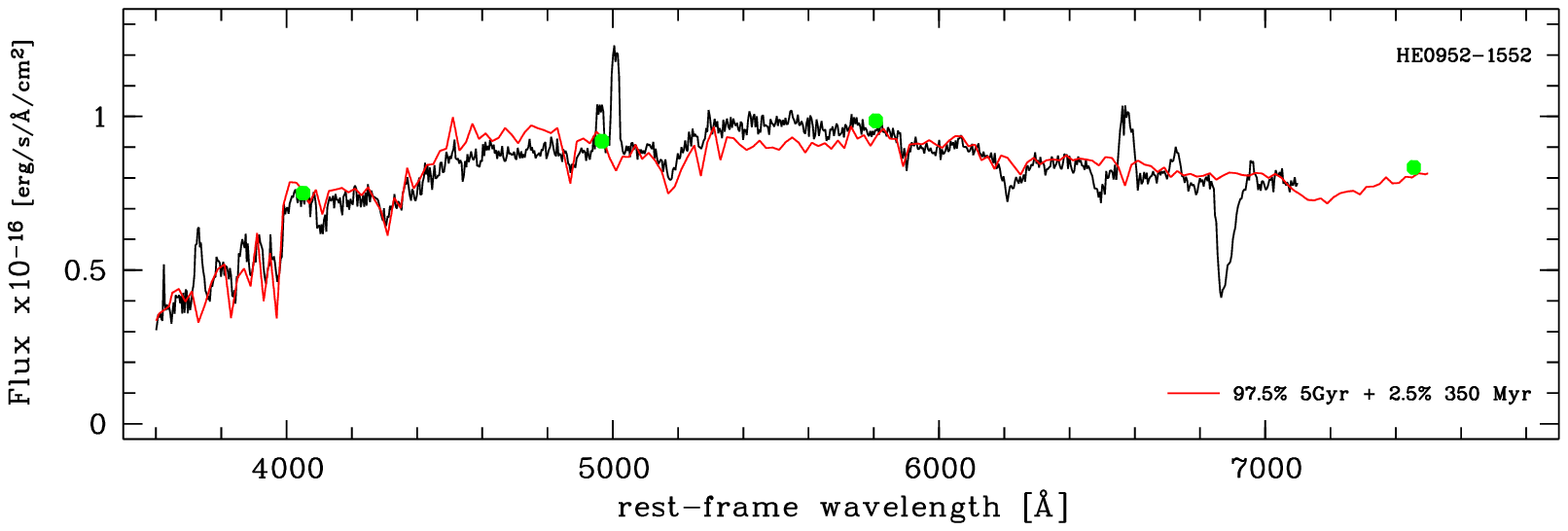}
\includegraphics[bb = 65 626 535 772,clip,angle=0,width=\plotwidth]{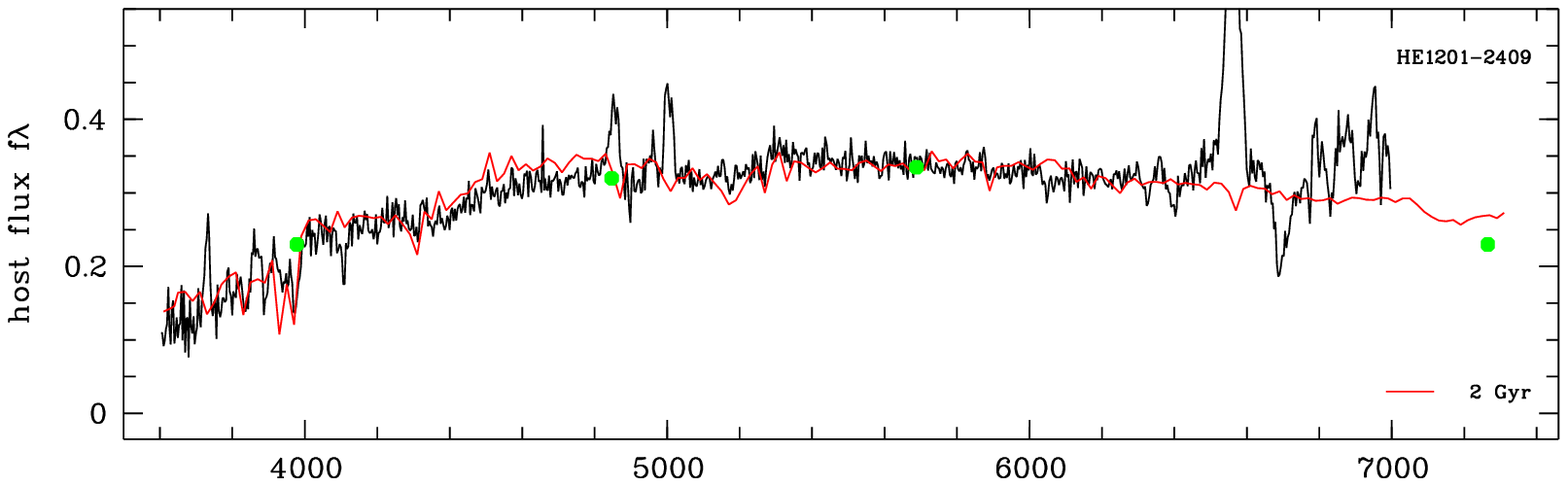}
\includegraphics[bb = 65 626 535 772,clip,angle=0,width=\plotwidth]{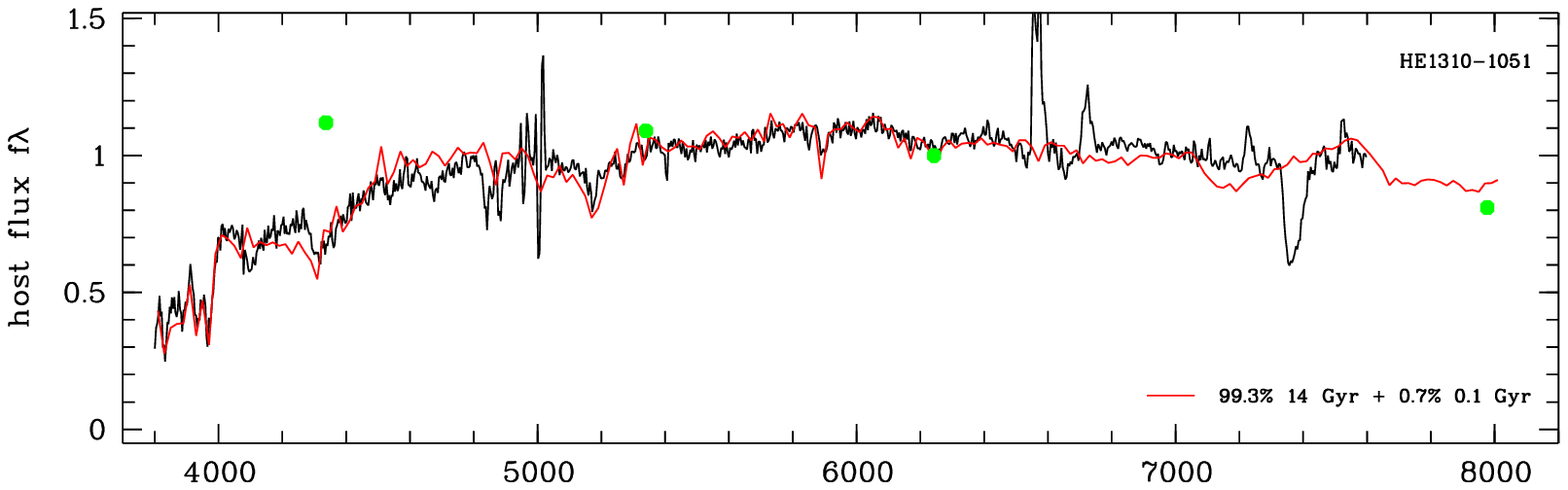}
\includegraphics[bb = 65 615 535 772,clip,angle=0,width=\plotwidth]{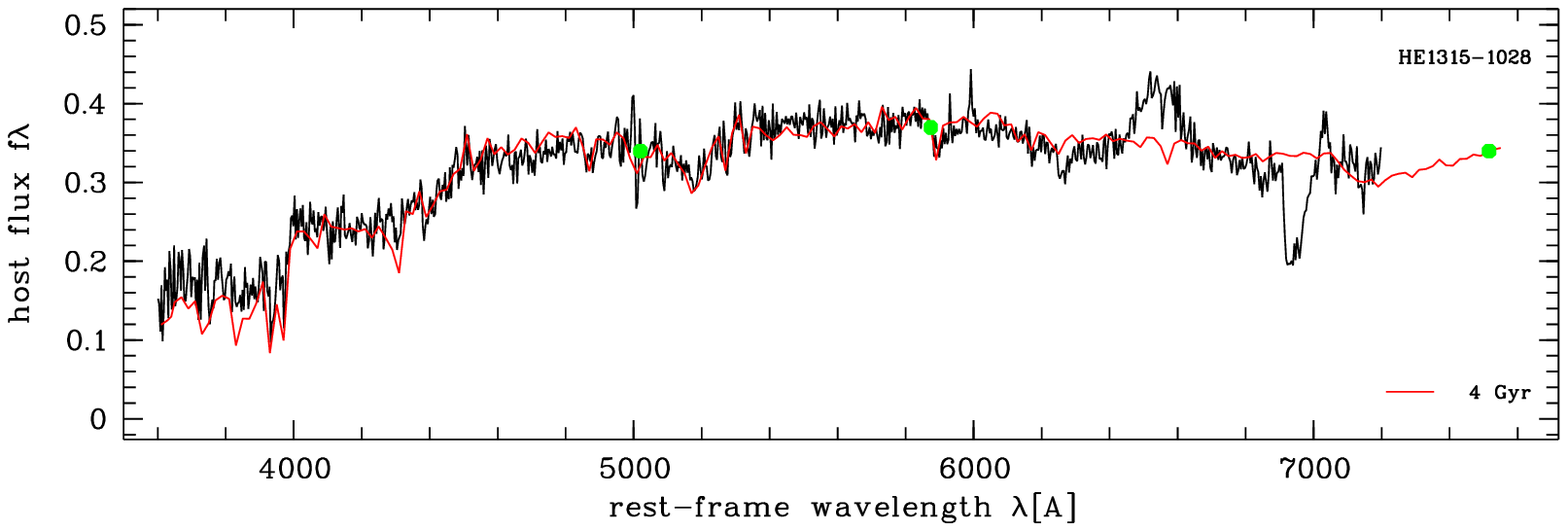}
\end{center}
\caption{\label{fig:efosc_sspfits} 
Model fits for four EFOSC objects with sufficient quality. See text for
description.
}
\end{figure*}

In this process we start by testing the absolute spectrophotometric
calibration of the data and precision of our decomposition algorith by
comparing the spectra to broad-band fluxes derived in \citep{jahn04a} for the
EFOSC sample. We convert the magnitudes in the $B$, $V$, $R$, and $i$ band to
fluxes using the appropriate filter curves and zero points. They are shown as
solid symbols in Fig.~\ref{fig:efosc_sspfits} overplotted over the host galaxy
spectra for four objects. By using a slight extrapolation of the spectrum to
the red by a stellar population model (see below) $V$, $R$, and $i$ are in
very good aggreement with the spectra.

For two objects this is also true for the $B$ band, it is missing for
HE\,1315--1028, and for HE\,1310--1051 $B$ it is off by a factor of 1.5. We
have no explanation for this. The $B$ band image for HE\,1310--1051 is of very
good quality, therefore the decomposition did not have large uncertainties. An
apparent asymmetry in the host galaxy, with potentially bluer colours due to
enhanced star formation after a merger was masked in the photometry and should
not boost the $B$ band flux over the spectrum flux, even if the slit has
missed this region. A fit to the optical and NIR colours of this host galaxy
suggests a higher fraction of a younger stellar population than the spectrum.
We take this as a hint that there might be an underestimation of the blue part
of the host galaxy spectrum in this case.
\smallskip

For both samples we we compare the host galaxy spectra to theoretical
evolution synthesis models, making use of the GALAXEV model family by
\citet{bruz03}. Used are models based on the Scalo IMF and of solar
metallicity. We used a grid of simple single stellar population (SSP) models
of different ages, from 0.01 to 14~Gyr, plus a continuous star formation (CSF)
model with constant star formation rate for 14~Gyr, created from the SSP
models. To this grid we added two SSP composite spectra. No internal dust
extinction in the galaxy was added.

\begin{figure*}
\begin{center}
\includegraphics[bb = 73 626 535 772,clip,angle=0,width=\plotwidth]{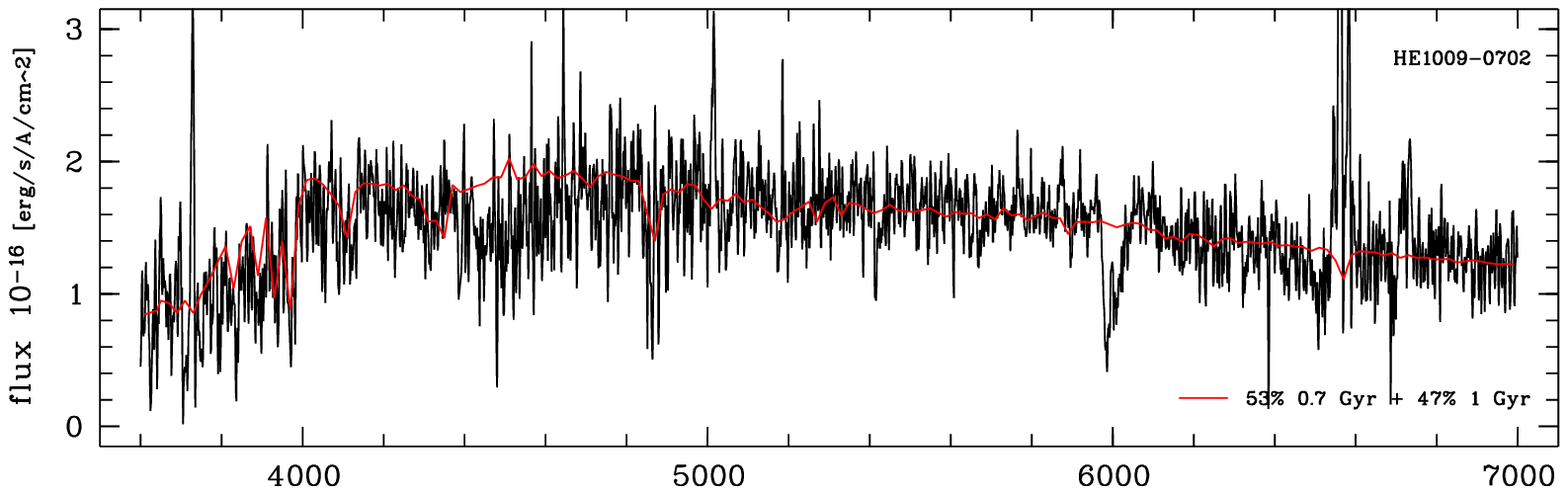}
\includegraphics[bb = 73 626 535 772,clip,angle=0,width=\plotwidth]{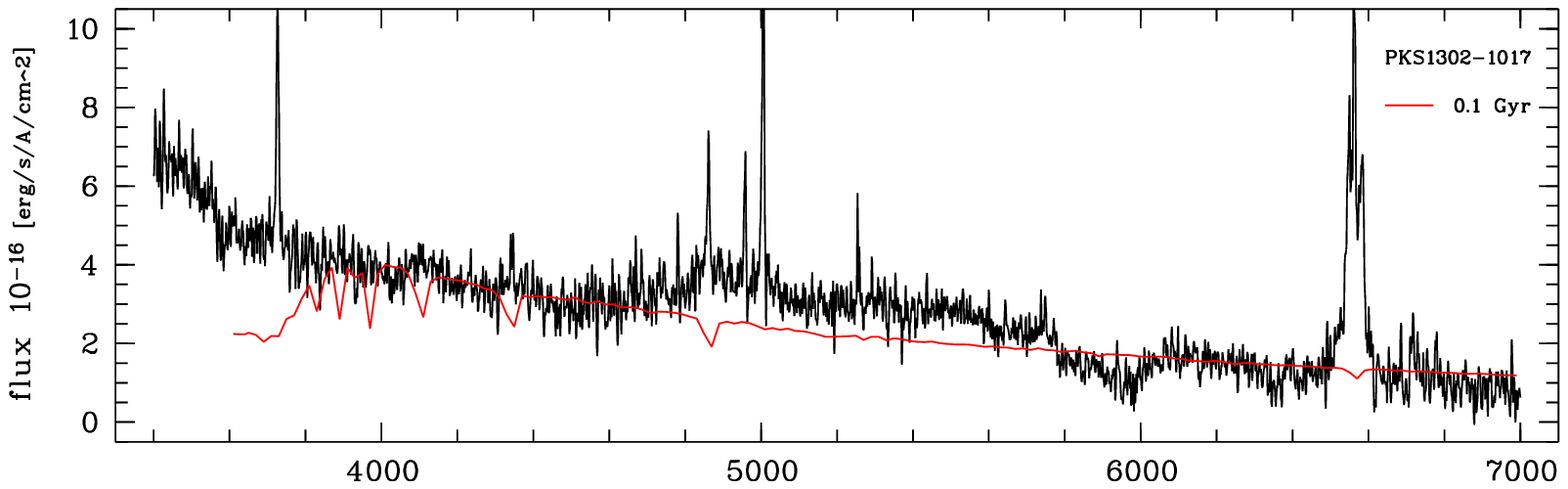}
\includegraphics[bb = 73 626 535 772,clip,angle=0,width=\plotwidth]{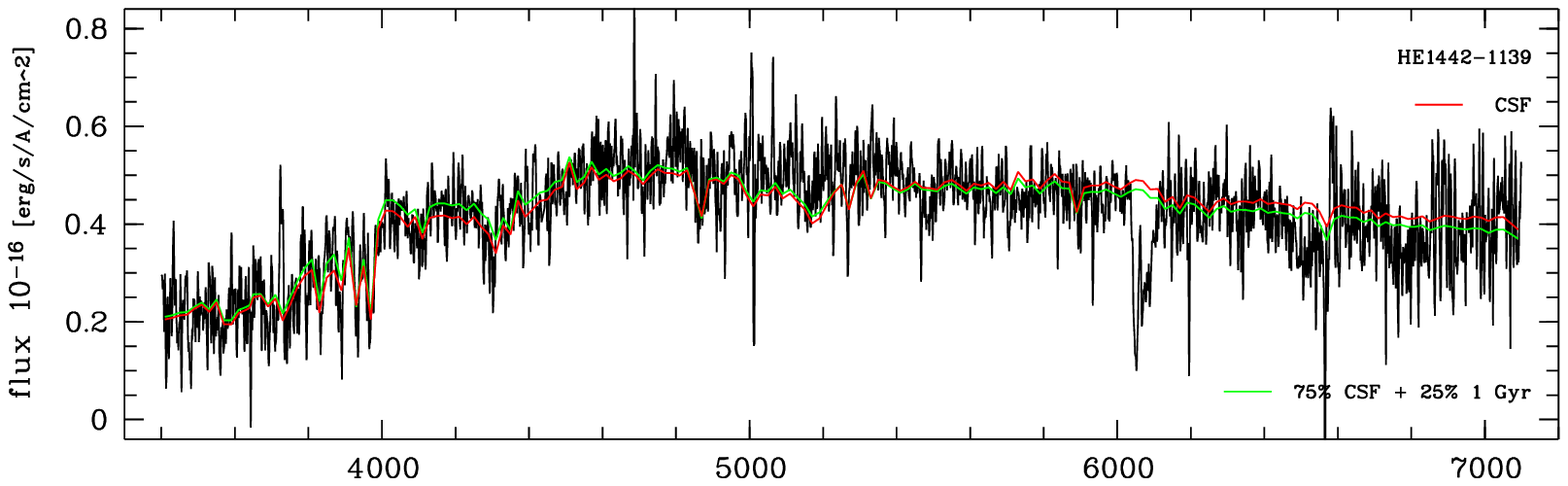}
\includegraphics[bb = 73 615 535 772,clip,angle=0,width=\plotwidth]{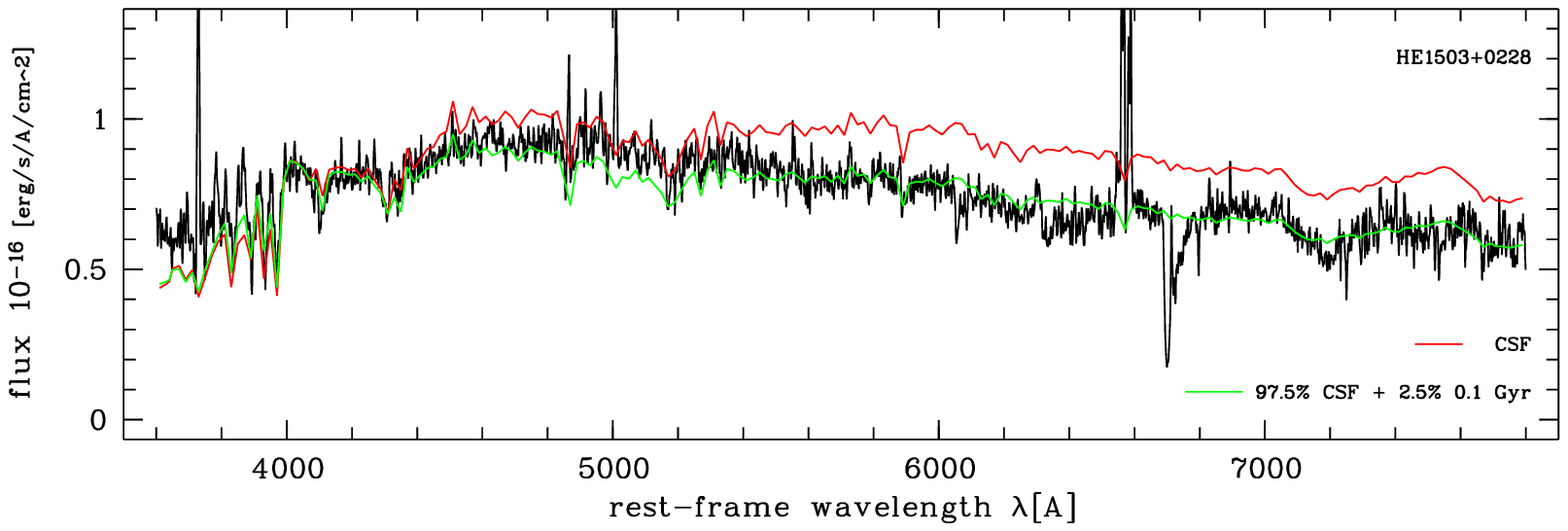}
\end{center}
\caption{\label{fig:vlt_sspfits} 
Spectral models for four VLT objects with good quality. See text for
description.
}
\end{figure*}

We did not perform a rigorous fit but tried to reproduce the spectral slope as
well as the absorption line strengths as good as possible with these
simplistic models. There are four sources each from the two samples where we
trust the decomposition to a degree that we attempt a fit to the overall SED
(Figures~\ref{fig:efosc_sspfits} and \ref{fig:vlt_sspfits}).

In the EFOSC sample (Fig.~\ref{fig:efosc_sspfits}) two models are roughly
matched by single stellar populations of 2~Gyr (HE\,1201--2409) and 4~Gyr
(HE\,1315--1028), however, for the latter the 4000\AA-break is slightly
smaller in the data than in the model, so it might be slightly younger. These
luminosity weighted ages are interesting. While for HE\,1315--1028 the age of
$\la$4~Gyr is rather typical, for the host is classified as disk dominated
\citep{jahn04a}, however HE\,1201--2409 was classified as a very compact early
type host galaxy, which would point to an excess of young stars from recent
star formation. From optical and near-infrared broad-band imaging we found an
age matching this result (1--2~Gyr).

The two other QSOs with EFOSC data can not be modelled with single stellar
populations. For HE\,0952--1552 we find that a combination of a stellar
population that is comparably old for a disk-dominated galaxy (5~Gyr) plus a
2.5\% contribution from a moderately young population (350~Myr) would work
best. The strength of the 4000\AA-break is slightly overestimated, pointing to
even more young stars, and there is a slight mismatch at 4500 and
5500\AA. However, the overall spectral slope is matched quite well.

HE\,1310--1051 requires only about 1\% of e.g.\ a 100~Myr population on top of
a very evolved population (14~Gyr). Here the 4000\AA-break is matched very
well. In total this could point to a quite old bulk of stars, with some recent
or ongoing star formation, while the morphological analysis classifies the
host as disk dominated. Broad band colour also favour a rather old average
stellar age of 6~Gyr, although the amount of young stars is higher due to the
leverage of the high $B$ band flux point.
\smallskip

In the FORS sample (Fig.~\ref{fig:vlt_sspfits}) HE\,1009--0702 is overall
rather young, with a mean stellar age of 0.7--1~Gyr, however the morphology is
ambiguous, suggesting a bulge plus disk (Tab.~\ref{tab:sampleprops}). Even
younger is the extreme case of PKS\,1302--1017, the only radio-loud quasar in
the sample. It has an elliptical host galaxy \citep{bahc97} as judged from HST
imaging, but there are indications for substantial deviation from a smooth
profile in ground based NIR imaging \citep{kuhl03}. The extracted spectra
deviate in the blue somewhat between our decomposition and the deconvolution
in \citet{leta06}, but the spectrum is consistently very blue, with an
extremely young stellar age. There are almost no discernable stellar
absorption lines, with only a slight CaK line, weak MgI and absent NaD. This
can not be artifacts of the decomposition since no broad emission components
of \ha and \hb are visible in the extracted host spectrum.

HE\,1442--1139 and HE\,1503+0228 are less surprising, being both
morphologically disk galaxies with very typical spectra. They are both best
fitted with continuous star formation models, plus an added 2.5\% of a 100~Myr
population in the case of HE\,1503+0228. Here also the absorption lines are
well matched, substantially better than for any EFOSC object.

\subsection{Peculiar velocities}\label{sec:velocities}
\citet{leta06} find about 50\% of the QSO sample to show signs of asymmetries
in the hosts or substantial companions near or even inside the host galaxy,
which is consistent with the fraction for the EFOSC sample \citep{jahn04a}. In
the FORS sample 12/20 hosts have sufficient S/N to determine spatially
resolved velocity information, and in six of these cases the velocity fields
are consistent with proper rotation curves. The remaining six cases show
either very little peculiar velocities (PKS\,1302--1017), or show strong signs
of interactions, with measured velocities (one side versus systematic
velocity) of up to several 100~km/s. The peculiar velocities for the FORS
sample are all consistent with the values found by \citet{leta06}.

In the EFOSC sample, data quality only permits to extract velocity information
for five of the eight objects in the sample -- we have stated the measurements
for HE\,0952--1552 already in Section~\ref{sec:0952}. While for the FORS
sample normally full rotation curves can be extracted, here the S/N is not
large enough to determine more than one velocity value for each side of the
host galaxy. Table~\ref{tab:efosc_rotation} summarizes the measured and
deprojected velocities for both samples.

\begin{table*}
\caption{\label{tab:efosc_rotation}
Geometrical, velocity, and stellar population properties of the EFOSC
sample. The columns list morphological type (as in
Table~\ref{tab:sampleprops}), inclination $i$, angle between slit and major
axis $\phi$, detection of rotation in gas emission lines, maximum velocities
of the gas as observed ($v_\mathrm{g,obs}$) and deprojected
($v_\mathrm{g,rot}$), and the same for stellar absorption lines
($v_\mathrm{s,obs}$, $v_\mathrm{s,rot}$). Velocities are in units of km/s. The
last column gives the single-stellar population fit ages from one or two
components, ``CSF'' stands for a continuous star formation model. A ``?'' is
given, if this property is not known, ``--'' if this property is not
accessible. For HE\,1315--1028 there are indications for a merging companion
in the slit. See text for details.
}
\begin{center}
\begin{tabular}{llccccccccc}
Object
&\multicolumn{1}{c}{Type}
&\multicolumn{1}{c}{$i$ [$^\circ$]}
&\multicolumn{1}{c}{$\phi$ [$^\circ$]}
&\multicolumn{1}{c}{Gas rot.}
&\multicolumn{1}{c}{$v_\mathrm{g,obs}$}
&\multicolumn{1}{c}{$v_\mathrm{g,rot}$}
&\multicolumn{1}{c}{Stell. rot.}
&\multicolumn{1}{c}{$v_\mathrm{s,obs}$}
&\multicolumn{1}{c}{$v_\mathrm{s,rot}$}
&\multicolumn{1}{c}{$t_\mathrm{stellar}$}\\
\hline
HE\,0952--1552& D &48&25&y&350&575&y&200&280&97.5\% 5\,Gyr\,+\,2.5\% 350\,Myr\\
HE\,1029--1401& E &--&--&y&204&-- &n&0&--&--\\
HE\,1201--2409& E &--&--&?&--&--&?&--&--&2\,Gyr\\
HE\,1239--2426& D &38&40&?&$<55$&$<140$&?&--&--&--\\
HE\,1310--1051& D &27&10&y&470&$>1000$&?&--&--&99.3\% 14\,Gyr\,+\,0.7\%\,100
Myr\\
HE\,1315--1028& D &55&50&?&--&--&y&$\sim$215&610?&4\,Gyr\\
\hline
HE\,0914--0031 &D &--&--&y&180  &--&--&--&--&--\\
HE\,1009--0702&E/D&--&--&y& 75  &--&?&--&--&0.7--1\,Gyr\\
HE\,1029--1401 &E &--&--&y&180  &-- &n&0 &--&--\\
PKS\,1302--1017&E &--&--&?&$<$40&-- &?&--&--&$<$100\,Myr\\
HE\,1434--1600 &E &--&--&y&140  &-- &?&--&--&--\\
HE\,1442--1139 &D &--&--&y&115  &--&y&$\sim$40&--&CSF\\
HE\,1503+0228  &D &43&28&y&160  &260&y&150&245&97.5\%\,CSF\,+\,2.5\%\,100\,Myr\\
\hline
\end{tabular}
\end{center}
\end{table*}

As noted, for HE\,0952--1552 the stellar velocities are consistent with
rotation of a massive disk galaxy, while the gas velocities require an
additional velocity component. Together with a spatial asymmetry of the
emission line (Fig.~\ref{fig:he0952_oiii}) this is potentially a sign for an
outflow.

For HE\,1029--1401 the decomposition was not completely successful in the
EFOSC data. The SED was unrealistic and much smaller residual are seen in the
FORS data (this is the only target common to both datasets). However, around
individual absorption and emission lines the decomposition residuals were
small, and enough to determine velocities. We measure $\sim$200~km/s
rotational velocity in \oiii, consistent with 180~km/s in the FORS data set,
and values consistent with zero in the CaK absorption line. So the stellar
velocities are consistent with the overall elliptical morphology of the host
galaxy, while there is either a disk of gas present in the host, or gas flows
in or out from opposite directions with 200~km/s projected velocity. The
``rotation curve'' from the FORS data is in fact not completely consistent
with rotation (rather a curve of constant slope than a classical rotation
curve), with either decomposition technique. If not decomposition residuals
are responsible for a skewing of the rotation curve, this is a clear sign that
the gas motion is dominated by other sources, flows or interaction.

HE\,1239--2426 is a ring galaxy, showing a strong ring of stars around a
central bulge or bar, the ring inclined by $\sim$40$^\circ$ (asuming
intrinsically circular geometry). We find only upper limits in the gas
velocities, so either the gas is not moving with more than 140~km/s or the
line of sight to the galaxy is in reality more face-on than estimated.

The two remaining host galaxies show peculiarities in their
velocities. HE\,1310--1051 allows to measure \oiii velocities of
470~km/s. With a fitted inclination of 27$^\circ$ and an angle of 10$^\circ$
between slit and major axis, the deprojected velocity is $>$1000~km/s. Since
the geometry of HE\,1310--1051 is severely distorted \citep{jahn04a}, both
angles have a substantial uncertainty. But even the observed velocity lies
outside the limits for disk rotation, so we trace a gas flow outside of a
dynamic equilibrium. The absorption lines have too low S/N to be studied.

For HE\,1315--1028 on the other hand the stellar absorption lines show signs
of peculiar velocities, while the host galaxy's emission lines are too weak to
have velocities measured. The stellar line of sight velocity is $\sim$160~km/s
in CaK and $\sim$275~km/s in NaD. The host galaxy has in principle a simple
geometry, with inclination of 50--60$^\circ$ and a slit angle of
$\sim$50$^\circ$. With these angles the projection is a factor of 2.8, which
would translate into a deprojected velocity of 610~km/s, clearly too much for
normal disk rotation. Since it seems unlikely that the inclination and major
axis positions are so far off to in fact reach a physically sensible
deprojected velocity, it is possible that we measure the velocity of a merging
companion. HE\,1315--1028 has a companion within its disk (in projection?)
that is covered by the slit. It is conceivable that this companion is moving
towards HE\,1315--1028 at a velocity comparible to the rotation velocity of
HE\,1315--1028, creating the projected velocity without stellar rotation of
this velocity.

 \subsection{Spatial distribution of the ionized ISM}\label{sec:ism_distribution}
 %

 \begin{figure*}
 \includegraphics[bb = 65 469 402 774,clip,angle=0,width=5.8cm]{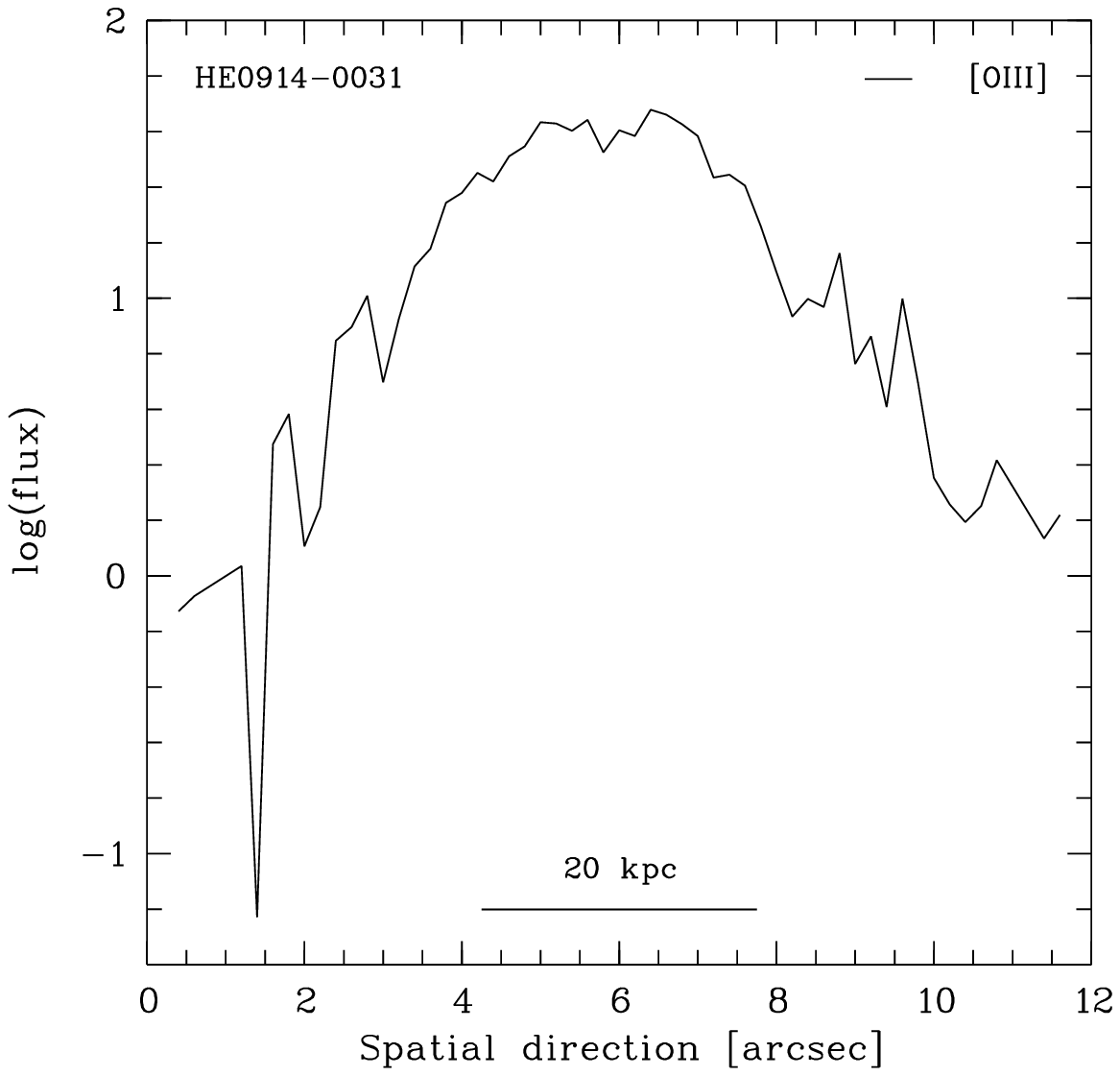}\hfill
 \includegraphics[bb = 65 469 402 774,clip,angle=0,width=5.8cm]{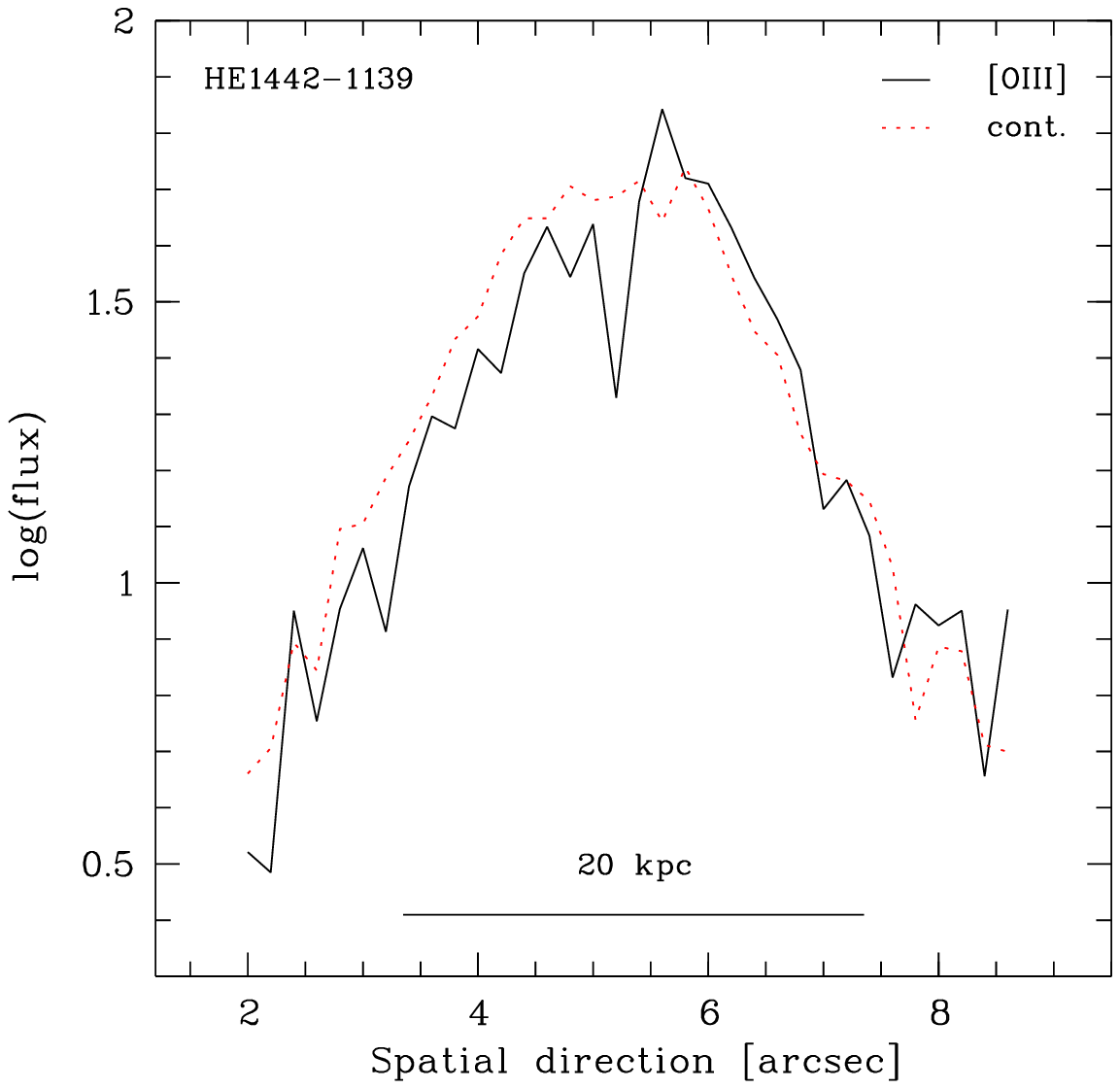}\hfill
 \includegraphics[bb = 65 469 402 774,clip,angle=0,width=5.8cm]{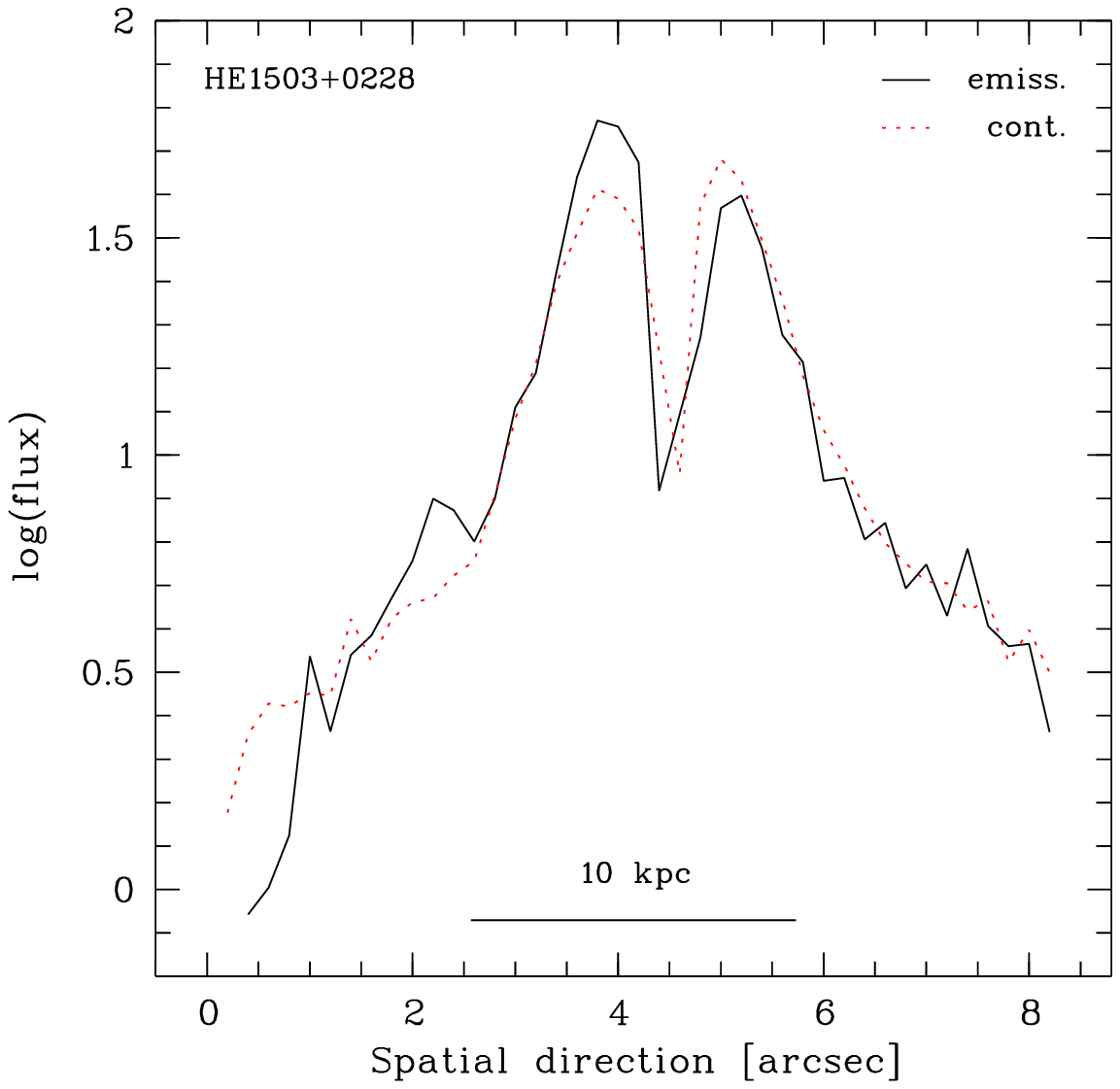}\\
 \includegraphics[bb = 65 453 402 774,clip,angle=0,width=5.8cm]{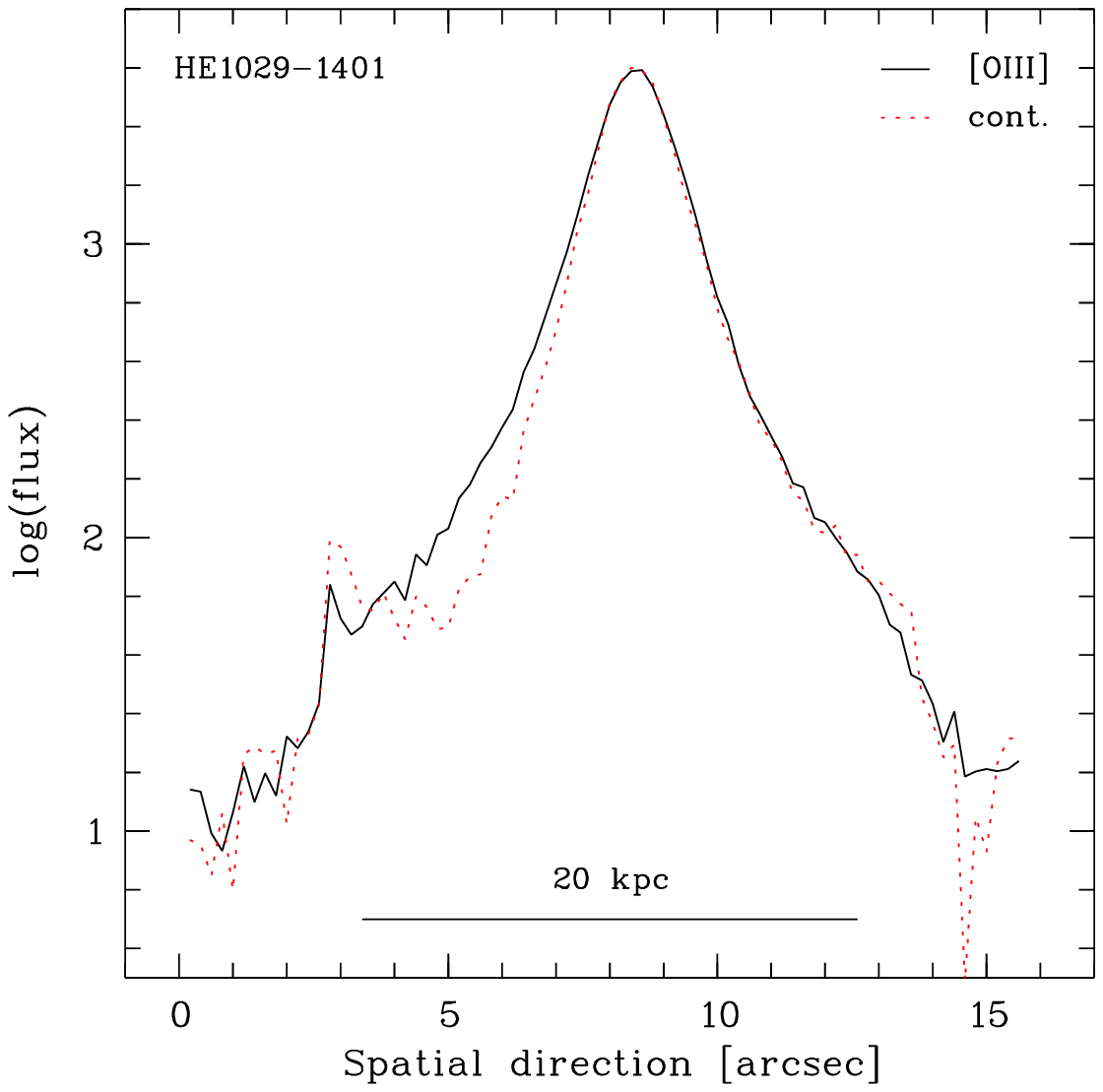}\hfill
 \includegraphics[bb = 65 453 402 774,clip,angle=0,width=5.8cm]{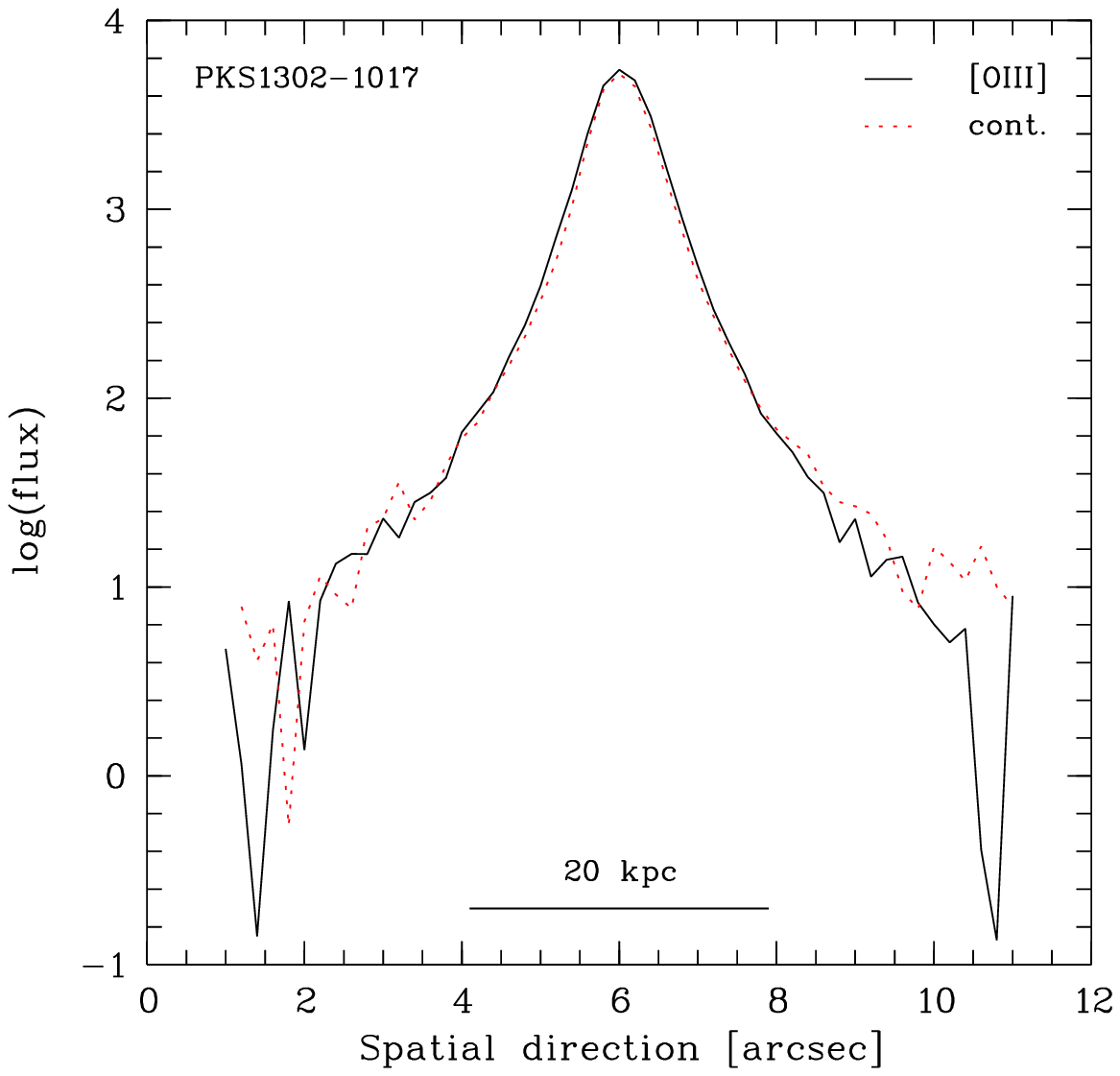}\hfill
 \includegraphics[bb = 65 453 402 774,clip,angle=0,width=5.8cm]{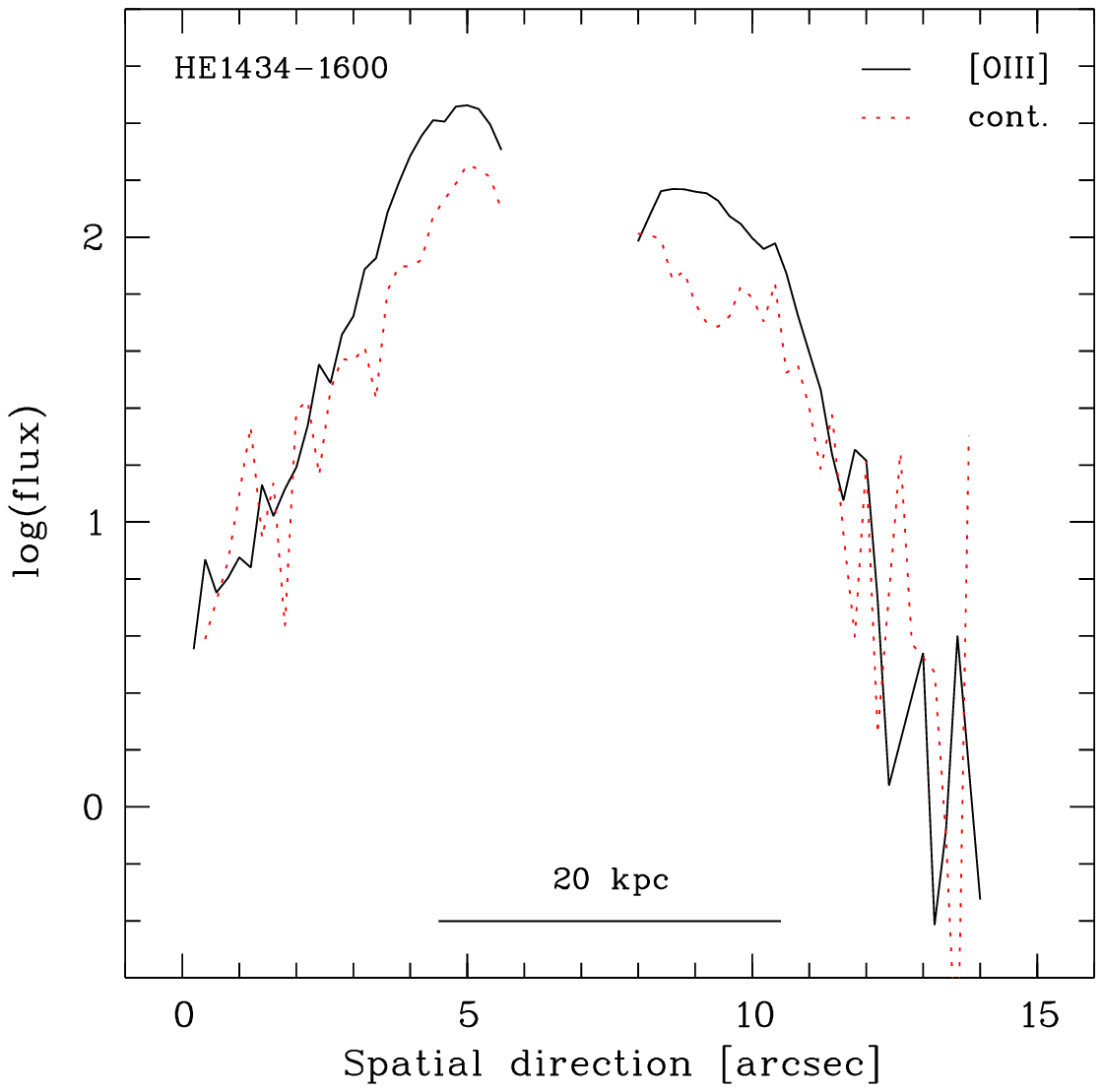}
 \caption{\label{fig:radial_profiles} 
 Radial surface brightness profiles of the \oiii 5007~\AA\ emission line and
 adjacent continuum. The solid curve is the emission line, the dashed red curve
 the continuum, scaled to match the emission in flux. For HE\,1503+0228 all
 emission lines in the $B$ grism were coadded.
 }
 \end{figure*}

 In the host galaxy spectra the spatial information from the slits is still
 available. After separation and special treatment of the narrow emission lines
 from the host galaxies (see Section~\ref{sec:emissionlines}) it is not only
 possible to measure line ratios to determine ionisation properties of the ISM
 as a whole, but also the spatial distribution of line emission can be compared
 to the stellar continuum emission. This can be complementary to an analysis of
 peculiar velocities to search for signs of irregularities. 

 In case that the ionising radiation reaches the outer limits of the host, and
 the ionising spectrum does not change, the radial profile of an emission line
 simply maps the distribution of the ISM. Thus if an emission line can be
 continuously traced out to the noise level, and if this coincides with the
 traceability of the continuum, then this is suggestive of ionisation
 throughout the galaxy. If the profiles also coincide, the distribution of
 stars and ISM is very similar (on the scales samples). If, however, at some
 radius the profile ends, one of two causes are possible. Either a limited
 extention of the gas, or, if the nucleus or a circumnuclear starburst were the
 source for ionised gas, and the number of high energy photons were limited so
 that only inside a given radius the gas was ionised.

 We show the logarithmic radial emission profiles of \oiii\ for 6 objects from
 the FORS sample in Figure~\ref{fig:radial_profiles}. They were produced as
 variance weighed averages of the spatial profile of \oiii\ (both components
 added), subtracting the underlying continuum and compare it to the shape of
 the adjacent continuum itself. For HE\,1503+0228 we coadded all available
 lines in the $B$ grism. For HE\,0914--0031 no continuum can be extracted,
 because this is not resolved. We ordered the objects by morphology as
 determined from imaging in \citet{jahn04a}, with three disks in the top row,
 the spheroidal dominated hosts in the bottom. 

 The disks more or less conform with the exponential disk profile, which would
 appear as a straight line in the log diagram, modified by the seeing:
 HE\,0914--0031 shows no cutoff or deviation from a pure exponential disk
 profile. For HE\,1442--1139 continuum and line emission are consistent, if one
 accounts for a slight shift in centers introduced by a sky line residual. The
 profiles of HE\,1503+0228 are disturbed in the centre as a result of
 modelling. While for the very centre the profiles are oversubtracted, they
 show an excess at around 1\farcs5 distance from the center, creating an
 inflection further out. This is a signature of the residuals of the nuclear
 model, not a sign for a spheroidal host. Line and continuum are compatible.

 While PKS\,1302--1017 shows two very similar profiles, supporting the case of
 no rotational velocity field detected in the line emission
 \citep[see][]{leta06}, the two other spheroid dominated early type hosts are
 more interesting.

 For HE\,1029--1401 peculiar gas motion is detected consistent with rotation
 of 150--200~km/s, despite that we find a clear spheroidal morphology and no
 trace of rotational velocities the stellar absorption lines. The right hand
 side of the line profile coincides with the profile of the continuum. While
 the central part reflects the PSF-broadened emision from the
 narrow-line-region of the AGN, the line emission on the left hand side
 significantly deviates from the continuum emission. Outside of a few kpc from
 the centre (outside the inflection points) the line emission is very similar
 to an exponential disk profile, while between 1 and 5~kpc distance from the
 centre the continuum emission clearly is not (the secondary peak at 3\farcs5
 is due to structure in the CCD and flatfielding and visible for all of the
 spectrum). This gives further support to the notion that the velocity seen
 for the ISM is actually rotational in nature and not an
 outflow. HE\,1029--1401 has likely a low mass stellar and gas disk that can
 only be seen in very deep imaging, but that shows up in the bright emission
 lines.

 HE\,1434--1600 is a very interesting case and has extensively been described
 in \citet{leta04}. In the spectrum, the central pixels are disturbed by
 modeling residuals. When scaling the continuum so that the outer parts of the
 two profiles coincide, an excess of line emission can be seen further in,
 i.e., the gas emission surface brightness drops faster with increasing radius
 than the continuum emission. In addition, the emission decreases again in the
 central part, so that close to the nucleus the line emission is less than
 further out \citep[this result is not affected by modelling residuals, see
 also][]{leta04}. While the nucleus is clearly the source of ionising
 radiation as judged from emission line rations, the gas does not extend all
 through the galaxy but is depleted in the centre, the system is not
 relaxed -- consistent with the result by \citet{leta04} that HE\,1434--1600
 is a potential post merging system.

 This drop in \oiii, might also be present in the stellar continuum, but is at
 least substantially weaker. The quality of the separation process is not
 sufficient to resolve this. However, the turnover is actually also the likely
 source for the modelling residuals: The model assumption of a smoothly
 decreasing galaxy profile is violated, thus both host and nuclear model are
 somewhat mismatched. This then is the result of the turnover, not its
 cause. What remains to be resolved is if the velocities and gas depletion are
 the result of the system merging with its nearest companion, if the AGN is
 blowing out gas on a kiloparsec scale, or if the gas is following a
 coordinated rotation in a disk around the bulge.

 \section{Discussion and summary}\label{sec:discussion}
 As we have shown, separating host galaxy and nuclear emission in on-nucleus
 spectra is possible. This works not only for low luminosity Seyfert nuclei,
 but also for nuclei substantially brighter than their host when integrated
 over the area of the slit. The limiting factor in the success and quality of
 the separation is clearly the quality of fixing the spatial light
 distributions of the components, i.e.\ the error in determining the PSF and
 the morphology of the host. A strong advantage, for higher nucleus--host
 contrasts even a requirement, is the incorporation of external information on
 the host galaxy's morphology from (comparably) cheap imaging data.

 We have shown for three cases that broad band host galaxy fluxes are reliably
 matched, showing that even spectrophotometric measurements can be performed
 under good conditions. Deconvolving the spatial domain at each wavelength
 numerically as done in \citet{cour00} delivers a very similar results, to
 consistently $<$10\% difference to our approach, while applying a completely
 complementary approach. As a difference to the spectral eigenvector
 decomposition technique described by \citet{vand06} -- yet again complementary
 -- both these methods retain the two-dimensional information of the spectrum,
 allowing an analysis of spatial structure.

 Any of the two decomposition/deconvolution methods is difficult to calibrate
 against the influence of PSF uncertainties or S/N issues by itself. Below a
 certain S/N both methods will break down, the exact point depending on the
 precision of PSF determination and the algorithmic details. While in the high
 S/N regime both methods work without large differences and errors, it is very
 valuable to actually have two completely independent methods that can both be
 applied in lower S/N cases to cross-check the results. Since lower S/N data
 are i) less expensive and easier to obtain, and ii) the only option for higher
 redshifts, having two independent decomposition techniques is extremely
 valuable.
 \smallskip

 Scientifically, both investigated samples show similar characteristics, which
 is not surprising, as their selection functions, redshifts, and magnitudes
 are similar. In both the FORS and EFOSC samples, about 50\% of the sources
 show signs of deviation from a symmetric and undisturbed morphology, as
 judged from their images \citep{kuhl03,jahn04a}, pointing to recent or
 ongoing events of interaction or merging with neighbors. Both samples have
 QSOs with bulge or disk dominated host galaxies, and for both samples
 luminosity weighted stellar ages are found fot most galaxies. While these are
 consistent with optical morphology for the disk dominated hosts, these are
 unusual for the bulge dominated ones. Aside from PKS\,1302--1017 these
 stellar populations, however, do not appear to be general starburst
 populations, but mixes of old and young population, or, if expressed as a
 single age, 1--2~Gyr of age. This is typical rather for intermediate to late
 type spirals, in clear contradiction to the observed morphology, if these
 were normal, average field galaxies.

 Interestingly enough, there are signs for systematic gas velocities found in
 both samples, not only in the disk dominated hosts, but also in the bulge
 dominated. On the other hand it is not clear, whether the source for this is
 simple disk rotation, e.g.\ that HE\,1029--1401 and HE\,1434--1600 are not
 consisting solely of a spheroid component, but have a comparably low
 luminosity stellar and gas disk as well. However, the velocity fields of the
 two systems just mentioned show deviations from a simple disk rotation
 \citep{leta04,leta06} law, which could also allow the interpretation that
 these might be signs of in- or outflows of ISM gas.

 The connection of a substantial distortion rate, signs for recent star
 formation and AGN activity is very suggestive and the earlier conclusions from
 imaging \citep{jahn04a} are reinforced by the presented spectra. Comparing
 broad band colours and host galaxy spectra to synthetic models both point, in
 connection with distorted velocity fields, to a interaction--AGN connection
 acting on these galaxies.

 In the light of scenarios of AGN feedback \citep[e.g.,][]{hopk06a} as a crucial
 factor in the creation of the blue cloud and red sequence bimodality in galaxy
 colour-magnitude diagrams, there are two questions to our datasets: First, do
 we see merging in connection with every bright quasar? Here the answer is no,
 not with the current data. There are several objects that have no discernable
 close companions or even signs of ongoing merger. Several of our sample
 objects are seemingly undistorted disk galaxies. However, seeing merger signs
 is an issue of spatial resolution. Recently Canalizo et al.\ (in prep)
 reobserved early type quasar host galaxies from the samples by \citet{dunl03},
 now with five orbits of HST ACS time each, instead of one orbit of
 WFPC2. While these hosts looked smooth and symmetric before in shallow data,
 they now show shells and whisps that mark these as merger remnants.

 The second question is, if we find traces for outflows, powered by the AGN,
 that will eventially evacuate the host galaxy of any ISM to truncate star
 formation. We see line emission, so signs of gas, in almost all
 hosts. Sometimes ionized by the nuclear spectrum, in the other half of the
 cases by hot stars. Interesting is the drop in emission line intensity in the
 centre of HE\,1434--1600, combined with a velocity field that suggests
 deviations from a normal Keplerian rotation. These can be counted as signs of
 interaction, but if the ISM in the central part of the galaxy has been blown
 out is difficult to support -- the data is more pointing towards interaction
 than AGN feedback.

 Since also none of the other objects shows large-scale signs of outflows, but
 either Keplerian rotation curves or coherent morphological and velocity field
 distortions, we see no indications for large scale flows in these data. If the
 evacuation of the galaxies' ISM is quenching the QSO activity very shortly
 after the blowout, this result is in accordance with the model, since these
 black holes are still strongly accreting. A connection of substantial merging
 and AGN activity, however, is supported for about half of the objects.

 \section*{Acknowledgments}
 KJ gratefully acknowledged support by the Studienstiftung des deutschen
 Volkes. This work was also supported by the DFG under grants Wi~1369/5--1
 and Schi~536/3-1. A very special Thank You goes to B.~Kuhlbrodt for
 fundamental discussions, happy debugging, and serious cooking.

\label{lastpage}

\end{document}